%% file: elsarticle-template-rev.tex
\documentclass[preprint,1pt]{elsarticle}
\input{preamble.tex}
\usepackage[margin=0.7in]{geometry}

\begin{document}

\begin{frontmatter}

\title{On Scaling Robust Feedback Control and State Estimation Problems \\ in Power Networks\tnoteref{label1}}
\tnotetext[label1]{This work is supported by the National Science Foundation under Grants ECCS 2151571 and CMMI 2152450.}

\author[inst1]{MirSaleh Bahavarnia\corref{cor1}}\ead{mirsaleh.bahavarnia@vanderbilt.edu}

\author[inst1]{Muhammad Nadeem}\ead{muhammad.nadeem@vanderbilt.edu}

\author[inst1,inst2]{Ahmad F. Taha}\ead{ahmad.taha@vanderbilt.edu}

\cortext[cor1]{The corresponding author is M.~Bahavarnia. Tel. +1-(615)-423-8936.}

\affiliation[inst1]{organization={Department of Civil and Environmental Engineering, Vanderbilt University},%Department and Organization
            addressline={2201 West End Avenue}, 
            city={Nashville},
            postcode={37235}, 
            state={TN},
            country={USA}}

\affiliation[inst2]{organization={Department of Electrical and Computer Engineering, Vanderbilt University},%Department and Organization
            addressline={2201 West End Avenue}, 
            city={Nashville},
            postcode={37235}, 
            state={TN},
            country={USA}}

\begin{abstract}
Many mainstream robust \textcolor{black}{control/estimation} algorithms for power networks are designed using the Lyapunov theory \textcolor{black}{as it} provides performance guarantees for \textcolor{black}{linear/nonlinear} models of uncertain power networks but comes at the expense of scalability and sensitivity. In particular, Lyapunov-based approaches rely on forming semi-definite programs (SDPs) that are \textit{(i)} not scalable and \textit{(ii)} extremely sensitive to the choice of the bounding scalar that ensures the strict feasibility of \textcolor{black}{the linear matrix inequalities (LMIs)}. This paper addresses these two issues \textcolor{black}{by employing a celebrated non-Lyapunov approach (NLA) from the control theory literature}. In lieu of linearized models of power grids, we focus on (the more representative) nonlinear differential algebraic equation (DAE) models and showcase the simplicity, scalability, and parameter-resiliency of NLA. For some power systems, the approach is nearly fifty times faster than solving SDPs via standard solvers with \textcolor{black}{almost} no impact on the performance. The case studies also demonstrate that NLA can be applied to more realistic scenarios in which \textit{(i)} only partial state data is available and \textit{(ii)} sparsity structures are imposed on the feedback gain. The paper also showcases that virtually no degradation in state estimation quality is experienced when applying NLA.
\end{abstract}


\end{frontmatter}

\section{Introduction}\label{sec:Intro}

In current and future power systems, many feedback control and state estimation algorithms will jointly be deployed to perform network-driven, real-time feedback control. \textcolor{black}{On one hand}, feedback control algorithms aim to robustly stabilize the frequency oscillations and improve the transient stability in the presence of unknown uncertainties \cite{kanchanaharuthai2005robust,chuang2016robust,nugroho2023load,nadeem2023on}. \textcolor{black}{On the other hand}, state estimation algorithms enable system operators to estimate the unmeasurable dynamic and algebraic states in the presence of unknown uncertainties and abrupt changes \cite{zhao2017dynamic,zhao2019theoretical,nadeem2022dynamic,nadeem2022robust}.

Many studies including the aforementioned robust control/estimation algorithms, have used Lyapunov theory, resulting in convex \textcolor{black}{SDPs}/LMI formulations. Lyapunov-based approaches are significantly powerful and have extensively been applied to design various robust controllers and estimators for diverse engineering applications \cite{boyd1994linear}. One of the main Lyapunov-based approaches has been built upon a well-known lemma, namely, Kalman-Popov-Yakubovich (KYP) Lemma (alternatively known as the \textit{Bounded Real Lemma}) \cite{boyd1994linear,gusev2006kalman}, to cast $\mathcal{H}_{\infty}$ control/estimation problems as convex SDPs via incorporating the LMIs. In power systems, such convex SDP/LMI formulations have widely been utilized to design robust $\mathcal{H}_{\infty}$ \textcolor{black}{control/estimation} algorithms. The central idea behind the $\mathcal{H}_{\infty}$ control/estimation \cite{zhou1998essentials} is that by minimizing the $\mathcal{H}_{\infty}$ norm, one minimizes the effects of the unknown uncertainties and abrupt changes on the desired performance of the control/estimation algorithm.   

In \cite{nadeem2023robust,nadeem2022dynamic,majumder2005lmi}, casting the control/estimation problems for DAE-modeled power systems as convex SDPs/LMI formulations via KYP Lemma, the authors propose dense robust $\mathcal{H}_{\infty}$ controllers and estimators, respectively. However, the Lyapunov-based control/estimation algorithms in the aforementioned research works are highly costly in terms of computational time. For instance, computing the Lyapunov-based dense $\mathcal{H}_{\infty}$ controller in \cite{nadeem2023robust} via MOSEK \cite{andersen2000mosek} SDP solver implemented in YALMIP \cite{lofberg2004yalmip} can take more than $4$ hours (on a personal computer). In the presence of abrupt model changes/uncertainties, it is not practically acceptable to utilize such highly time-consuming $\mathcal{H}_{\infty}$ syntheses. 

Although the Lyapunov-based approaches provide a solid theoretical framework to analyze the control/estimation problems in terms of stability and performance, similar to any other LMI-based approaches, they face two main issues, \textcolor{black}{namely,} computational scalability and computational sensitivity. 

The former issue occurs when a Lyapunov-based approach casts a control/estimation problem as a convex SDP via incorporating the LMIs. It is evident that dealing with large-scale LMIs and solving the corresponding SDPs are generally challenging tasks in terms of scalability \cite{majumdar2020recent}. The authors in \cite{majumdar2020recent} have surveyed some alternative scalable (yet possibly inaccurate) SDP solvers like SDPNAL+ \cite{yang2015sdpnal+,sun2020sdpnal+} which is particularly developed based on a majorized semi-smooth Newton-CG augmented Lagrangian method. Although SDPNAL+ has significantly improved the scalability in a class of robust traffic density estimation problems \cite{nugroho2021control}, it has unfortunately been unable to even propose a feasible solution for the $\mathcal{H}_{\infty}$ control/estimation problem of \textcolor{black}{a medium-size power system.}

The latter issue occurs due to the fine-tuning of $\epsilon I$-shifting (where $\epsilon$ is the bounding scalar that ensures the strict feasibility of the LMIs and $I$ is the identity matrix) in LMIs as there is no systematic way in the literature to appropriately choose/tune $\epsilon$ in general. Note that the appropriate choice of $\epsilon$ positively affects the feasibility and sub-optimality of the solution at the cost of excessive computational time. The $\epsilon I$-shifting idea is commonly utilized to implement the strict LMIs as the non-strict LMIs because SDP solvers cannot deal with the strict LMIs. 

One might think that utilizing the generalized algebraic Riccati equations (GAREs) could be effective \textcolor{black}{in overcoming the issues mentioned above}. Unfortunately, although casting the control/estimation problems for a DAE-modeled power system as GAREs is theoretically possible (thus increasing the scalability significantly)  \cite{takaba1994h,kawamoto1997standard,feng2017robust}, solving the cast GAREs practically faces numerical issues/difficulties due to checking a rank equality constraint and an indefinite matrix coefficient appearing in an algebraic quadratic equation (AQE). In fact, such an AQE is efficiently solvable if that matrix coefficient is not indefinite which is not the case for a DAE-modeled power system in general. Then, casting the $\mathcal{H}_{\infty}$ control/estimation problems as GAREs to bypass the aforementioned scalability and sensitivity issues in the case of DAE-modeled power systems, unfortunately, becomes impractical.

\textcolor{black}{To that end, in light of the aforementioned limitations and to effectively overcome the arising issues by the Lyapunov-based approaches (i.e., computational scalability and computational sensitivity), considering the NDAE state-space model of a power system, we derive the equivalent nonlinear ordinary differential equation (ENODE) state-space model without imposing any simplifying specific form on the controller or the estimator.} It is noteworthy that the continuous-time algebraic Riccati equation (CARE)-based approaches \textcolor{black}{are not} applicable to the derived ENODE state-space model as the dependency of the state-space matrices \textcolor{black}{on the controller or the estimator is not} affine (e.g., $A + BK$ or $A + LC$) unlike the simplified state-space model derived in \cite{nadeem2023robust}. To design $\mathcal{H}_{\infty}$ controllers and estimators for given NDAE state-space models, motivated by the aforementioned twofold issues, we \textcolor{black}{employ} a \textcolor{black}{celebrated} non-Lyapunov approach (NLA) \textcolor{black}{from the control theory literature \cite{apkarian2006nonsmooth}} and utilize off-the-shelf computational tools along with the derived ENODE state-space model to bypass the computational scalability and computational sensitivity issues arising in Lyapunov-based $\mathcal{H}_{\infty}$ \textcolor{black}{controller and estimator} designs for DAE-modeled power systems. \textcolor{black}{The proposed $\mathcal{H}_{\infty}$ synthesis (NLA) in \cite{apkarian2006nonsmooth} is mainly built upon non-convex non-smooth optimization techniques taking advantage of the Clarke sub-differential \cite{clarke1990optimization} computation of the $\mathcal{H}_{\infty}$ norm. Such a structured $\mathcal{H}_{\infty}$ synthesis has found extensive control applications including but not limited to proportional-integral-derivative (PID) control and combination of proportional-integral (PI) and SOF control of distillation column, aircraft autopilot, and helicopter as highlighted by \cite{gahinet2011structured,gahinet2011decentralized,gahinet2012frequency}.}

Note that NLA can alternatively be applied to any $\mathcal{H}_{\infty}$ \textcolor{black}{controller and estimator} design problem that can potentially be cast as a convex SDP/LMI formulation, e.g., \cite{werner2003robust,majumder2005lmi,kanchanaharuthai2005robust,nadeem2022dynamic,nadeem2023robust,basati2023robust}. Moreover, compared to such Lyapunov-based syntheses in the literature, we take into account a more general problem setup consisting of structured \textcolor{black}{controllers and estimators} instead of solely dense counterparts. In other words, thanks to the specialized features of NLA, we are able to propose structured $\mathcal{H}_{\infty}$ controllers and estimators.

The paper's contributions are as follows:
\begin{itemize}
\item We design structured $\mathcal{H}_{\infty}$ controllers and estimators for DAE-modeled power systems via \textcolor{black}{employing a celebrated NLA from the control theory literature \cite{apkarian2006nonsmooth}}. Specifically, for the controller design, the proposed controller is applicable to a general scenario in which we \textcolor{black}{deal with} partially-accessed states and noisy measurement outputs. Moreover, for the estimator design, the proposed estimator can deal with load demand and renewable disturbances, non-Gaussian measurement noise, and unknown control inputs.

\item NLA is noticeably faster than the Lyapunov-based counterpart while attaining the same $\mathcal{H}_{\infty}$ performance (computational scalability). Moreover, there is no need to deal with the fine-tuning of the bounding scalar that ensures the strict feasibility of the LMIs anymore as we have successfully bypassed it (computational sensitivity).

\item \textcolor{black}{We investigate the conditions that can potentially affect the superiority of NLA over its Lyapunov-based counterpart in terms of computational scalability. It turns out that having a relatively large number of inputs (for the controller design) and outputs (for the estimator design) can negatively affect such a superiority. Beyond a specific level, such superiority can even be reversed.}

\item Utilizing a notion of \textit{block-sparsity}, we study the impact of various sparsity structures on the $\mathcal{H}_{\infty}$ performance of the proposed controller design. Also, we empirically explore the potential effects of output matrix selection on the $\mathcal{H}_{\infty}$ performance and how we can take advantage of such empirical information to select an optimal output matrix. Specifically, we can detect the states that have a central role in the stabilization of the system. 

\end{itemize}

\noindent {\bf Notations.} Uppercase and lowercase letters denote matrices and vectors, respectively. We denote the set of real-valued $r$ by $c$ matrices by $\mathbb{R}^{r \times c}$. Also, we denote the $r$ by $c$ identity and all-ones matrices with $I_{r \times c}$ and $\mathbf{1}_{r \times c}$, respectively. To represent an $n$-dimensional identity matrix, we utilize $I_n$. For a matrix $M$, symbols $M^T$, $M^{\perp}$, $\bar{\sigma}[M]$, and $\|M\|_0$ denote its transpose, orthogonal complement, largest singular value, and number of non-zero elements, respectively. We denote positive semi-definiteness and positive/negative definiteness by $\succeq 0$ and $\succ 0$/$\prec 0$, respectively. For a square matrix $M$, we define $\mathbf{He}(M):= M + M^T$, and symbol $\mathbf{sa}(M)$ denotes the spectral abscissa of $M$, i.e., the maximum real part of the eigenvalues of $M$. Symbol $\ast$ represents the corresponding symmetric components of a symmetric matrix. We use $j$ and $\Re(s)$ to showcase the $\sqrt{-1}$ and the real part of complex-valued scalar $s$, respectively. We denote the matrix element-wise Hadamard product and the matrix Kronecker product with $\odot$ and $\otimes$, respectively. For a vector $v$, symbol $\|v\|$ denotes the Euclidean norm of $v$. Given a vector $h(t)$ defined for $t \in [0,\infty)$, its $\mathcal{L}_2$-norm is defined as $\sqrt{\int_0^\infty \|h(t)\|^2 dt}$. To save space, we omit the dependency on time, i.e., $(t)$ in denoting the time-dependent quantities. The set of subgradients of a function $f$ at the point $x$ is defined as the subdifferential of $f$ at $x$ and is denoted by $\partial f(x)$. We represent the function composition operator by $\circ$. The standard basis vector $\mathbf{e}_i$ represents a column vector with $1$ at the $i^{th}$ position and $0$ at the rest of the positions.

The rest of the paper is organized as follows: Section \ref{sec:Prel} consisting of preliminaries are three-fold: \textit{(i)} Section \ref{PSDAE} details the NDAE state-space model of the power system, \textit{(ii)} Section \ref{secAPF} elaborates on the problem statement, and \textit{(iii)} Section \ref{AR} briefly reviews dense $\mathcal{H}_{\infty}$ controller and estimator designs built upon the Lyapunov-based approach for DAE-modeled power systems. Section \ref{sec:Met} contains the central result of the paper which is the implementation of NLA to design structured $\mathcal{H}_{\infty}$ controllers and estimators for DAE-modeled power systems. Section \ref{sec:Example} via various benchmarks thoroughly verifies the effectiveness of NLA in comparison with its Lyapunov-based counterpart. Finally, Section \ref{Con} ends the paper with a few concluding remarks.

\section{Preliminaries} \label{sec:Prel}

In Section \ref{PSDAE}, we include the details of the NDAE state-space model of the power system. We formally state the structured $\mathcal{H}_{\infty}$ controller problem for DAE-modeled power systems in Section \ref{secAPF}. The $\mathcal{H}_{\infty}$ estimator problem can also be stated in a similar fashion. \textcolor{black}{Section \ref{AR} briefly sheds light on the Lyapunov-based approach utilized by \cite{nadeem2023robust,nadeem2022robust} to design dense $\mathcal{H}_{\infty}$ controllers and estimators for DAE-modeled power systems.}

\subsection{Power system NDAE state-space model} \label{PSDAE}

Here, we include the fourth-order dynamics of the modeled synchronous generators. \textcolor{black}{To this end}, we borrow the notations from \cite{nugroho2023load}. Let us consider a power system comprised of $N_b$ buses, modeled by a graph $(\mathcal{N},\mathcal{E})$ where $\mathcal{N}$ and $\mathcal{E}$ respectively denote the set of nodes and edges. The nodes are of three main types: \textit{(i)} traditional synchronous generators, \textit{(ii)} renewable energy resources (RERs), and \textit{(iii)} load buses. Note that $\mathcal{N} = \mathcal{G} \cup \mathcal{R} \cup \mathcal{L} \cup \mathcal{U}$ holds where $\mathcal{G}$, $\mathcal{R}$, $\mathcal{L}$, and $\mathcal{U}$ respectively collect $G$ generator buses, the buses containing $R$ renewables, $L$ load buses, and $U$ non-unit buses.

A fourth-order dynamics of synchronous generators can be modeled as \cite{nugroho2023load,taha2019robust,sauer2017power}
\begin{subequations} \label{eq:SynGen}
	\begin{align}
	\dot{\delta}_{i} &= \omega_{i} - \omega_{0}, \label{eq:SynGen1} \\ 
	\begin{split}
	M_{i}\dot{\omega}_{i} &= T_{\mr{M}i}-P_{\mr{G}i}- D_{i}(\omega_{i}-\omega_{0}), \end{split}\label{eq:SynGen2}    \\ 
	T'_{\mr{d0}i}\dot{E}'_{i} &= -\tfrac{x_{\mr{d}i}}{x'_{\mr{d}i}}E'_{i} +\tfrac{x_{\mr{d}i}-x'_{\mr{d}i}}{x'_{\mr{d}i}}v_i\cos(\delta_{i}-\theta_i) + E_{\mr{fd}i},  \label{eq:SynGen3} \\
	T_{\mr{CH}i}\dot{T}_{Mi} &= -T_{\mr{M}i} - \tfrac{1}{R_{\mr{D}i}}(\omega_{i}-\omega_{0}) + T_{\mr{r}i}, \label{eq:SynGen4}  
	\end{align} 
\end{subequations}
where $(\delta_i,\omega_i,E'_i,T_{Mi})$ and $(E_{fdi},T_{ri})$ respectively represent the generator's internal states and the generator's inputs. The generator's internal states, namely, $(\delta_i,\omega_i,E'_i,T_{Mi})$, its supplied power $(P_{Gi},Q_{Gi})$, and terminal voltage $v_i$ are related to each other via the following two algebraic equations \cite{nugroho2023load,taha2019robust}:  
\begin{subequations}\label{eq:SynGenPower}
	\begin{align}
		\begin{split}
		P_{\mr{G}i} &= \tfrac{\sin(\delta_i-\theta_i)}{x'_{\mr{d}i}} (E'_{i}v_i -\tfrac{(x_{\mr{q}i}-x'_{\mr{d}i}) v_i^2\cos(\delta_i-\theta_i)}{x_{\mr{q}i}}),
		\end{split}
	 \label{eq:SynGenPower1} \\
	\begin{split}
		Q_{\mr{G}i} &= \tfrac{E'_{i}v_i\cos(\delta_i-\theta_i) - v_i^2 \cos^2(\delta_i-\theta_i)}{x'_{\mr{d}i}}-\tfrac{v_i^2 \sin^2(\delta_i-\theta_i)}{x_{\mr{q}i}}.
	\end{split}\label{eq:SynGePower2}
	\end{align}
\end{subequations}
The power flow (PF)/balance equations governing the power transfer among generators, RERs, and loads are as follows \cite{nugroho2023load,sauer2017power}:
\begingroup
\allowdisplaybreaks 
\begin{subequations} \label{eq:GPF}
	\begin{align} 
	\begin{split}
\hspace{-0cm}P_{\mr{G}i} + P_{\mr{R}i}	+P_{\mr{L}i} \hspace{-0cm}&=\hspace{-0cm} \sum_{j=1}^{N_b}\hspace{-0.05cm} v_iv_j\hspace{-0cm}\left(G_{ij}\cos \theta_{ij} \hspace{-0cm}+ \hspace{-0cm}B_{ij}\sin \theta_{ij}\right),
\end{split}\label{eq:GPF1}\\
	\begin{split}
	\hspace{-0cm}Q_{\mr{G}i} + Q_{\mr{R}i}	+Q_{\mr{L}i}\hspace{-0cm} &=\hspace{-0cm} \sum_{j=1}^{N_b}\hspace{-0cm} v_iv_j\hspace{-0cm}\left(G_{ij}\sin \theta_{ij} \hspace{-0cm}- \hspace{-0cm}B_{ij}\cos \theta_{ij}\right),
\end{split}\label{eq:GPF2}		
	\end{align}
\end{subequations}
\endgroup 
where $\theta_{ij} := \theta_i - \theta_j$. The pair $(G_{ij},B_{ij})$ respectively denotes the conductance and susceptance between buses $i$ and $j$. Also, the pairs $(P_{Ri},Q_{Ri})$ and $(P_{Li},Q_{Li})$ respectively represent the active and reactive powers generated by the renewables and the active and reactive powers consumed by the loads.

Let us consider the following NDAE state-space model:
\begin{subequations}\label{NDAE}
\begin{align}
E \dot{x} &= A x + B u + B_w w + h(x,u,w),\label{NDAEEq1} \\
u &= F y,~ y = C_y x + D_y w, \label{eq:input}
\end{align} 
\end{subequations}
where $E$ denotes a singular matrix encoding the algebraic equations with all-zeros rows, $x = \begin{bmatrix} x_d^T & x_a^T \end{bmatrix}^T \in \mathbb{R}^{n_x}$, $x_d \in \mathbb{R}^{n_d}$, $x_a \in \mathbb{R}^{n_a}$, $u \in \mathbb{R}^{n_u}$, $w \in \mathbb{R}^{n_w}$, and $y \in \mathbb{R}^{n_y}$ respectively represent the state, dynamic state, algebraic state, control input, disturbance input, and measurement output vectors, $F$ denotes the static output feedback (SOF) controller matrix, and $A$, $B$, and $B_w$ are extracted via Jacobian-based linearization while $h(x,u,w)$ encompasses the linearization error vector associated with the nonlinearities. Here, we emphasize the point that we have \textbf{not} utilized a purely linearized model by overlooking the linearization error vector associated with the nonlinearities.   

To construct the NDAE state-space representation of the power system for the controller design, let us define
\begin{align*}
     x_d &= \begin{bmatrix}
        \delta^T & \omega^T & E'^T & T_M^T
    \end{bmatrix}^T,~
     x_a = \begin{bmatrix}
        a^T & \tilde{v}^T
    \end{bmatrix}^T,~a := \begin{bmatrix}
        P_G^T & Q_G^T,
    \end{bmatrix}^T,\\
    \tilde{v}&:= \begin{bmatrix}
        v^T & \theta^T
    \end{bmatrix}^T,~
    u = \begin{bmatrix}
        E_{fd}^T & T_r^T
    \end{bmatrix}^T,~ w = \begin{bmatrix}
        P_R^T & Q_R^T & P_L^T & Q_L^T
    \end{bmatrix}^T,
\end{align*}
where
\begin{align*}
    \delta &:= \{\delta_i\}_{i \in \mathcal{G}},~\omega := \{\omega_i\}_{i \in \mathcal{G}},~E' := \{E'_i\}_{i \in \mathcal{G}},~
     T_{M} := \{T_{Mi}\}_{i \in \mathcal{G}},\\
     P_{G} &:= \{P_{Gi}\}_{i \in \mathcal{G}},~Q_{G} := \{Q_{Gi}\}_{i \in \mathcal{G}},~v := \{v_i\}_{i \in \mathcal{N}},~\theta := \{\theta_i\}_{i \in \mathcal{N}},\\
     E_{fd} &:= \{E_{fdi}\}_{i \in \mathcal{G}},~T_{r} := \{T_{ri}\}_{i \in \mathcal{G}},~
     P_{R} := \{P_{Ri}\}_{i \in \mathcal{R}},~Q_{R} := \{Q_{Ri}\}_{i \in \mathcal{R}},\\ P_{L} &:= \{P_{Li}\}_{i \in \mathcal{L}},~Q_{L} := \{Q_{Li}\}_{i \in \mathcal{L}}.
\end{align*} 
Based on the above vector representation and \eqref{eq:SynGen}, \eqref{eq:SynGenPower}, and \eqref{eq:GPF}, the NDAE state-space model \eqref{NDAEEq1} for the power system can be constructed.

\subsection{Problem statement} \label{secAPF}

Assuming the $\mathcal{L}_2$-norm boundedness of $h(x,u,w)$ and considering it as $B_h w_h$, the NDAE state-space model \eqref{NDAE} reduces to the following linear DAE (LDAE) form:
\begin{subequations} \label{DAE}
\begin{align}
 E \dot{x} &= (A + BFC_y) x + \hat{B}_w \tilde{w},~\tilde{w} := \begin{bmatrix} w^T & w_h^T \end{bmatrix}^T,\\
\hat{B}_w &:= \begin{bmatrix} B_w + B F D_y & B_h \end{bmatrix}.
\end{align}
\end{subequations}With that in mind, to design an SOF $\mathcal{H}_\infty$ controller, we define the following performance output vector:
\begin{align}
     z &= C x + D u + \hat{D}_w \tilde{w},~ \hat{D}_w := \begin{bmatrix} D_w & 0 \end{bmatrix}. \label{zC0}
\end{align}
The main idea is to design an SOF $\mathcal{H}_\infty$ controller $F$ in \eqref{eq:input} such that it satisfies $\mathcal{H}_\infty$ performance criterion, \textcolor{black}{namely,} $\norm{z}^2_{\mathcal{L}_2} <  \mu^2\norm{{\tilde{w}}}^2_{\mathcal{L}_2}$ \cite{masubuchi1997h} ($\mu$: the $\mathcal{H}_\infty$ value). In the sequel, we elaborate on structurally-constrained $\mathcal{H}_\infty$ SOF controller design which frequently appears in more realistic scenarios. Further details about the design of such an SOF controller $F$ are later given in Section \ref{sec:Met}.

The sparsity structure of the SOF controller $F$ is of high significance \textcolor{black}{in decreasing} the communication burden among generators. To name a few common sparsity structures, we list the following two crucial sparsity structures: \textit{(i)} decentralized, and \textit{(ii)} distributed. In the context of power systems, each generator has access only to its own state information for a decentralized sparsity structure while in the case of a distributed sparsity structure, each generator has access to the state information of a few numbers of the other generators. To encode the sparsity structures, we simply use the matrix element-wise Hadamard product as follows:
\begin{align}
F &= F \odot S, \label{SS}
\end{align}
where $S \in \{0,1\}^{n_u \times n_y}$ represents an imposed binary sparsity structure. To that end, to design the structured SOF (SSOF) $\mathcal{H}_{\infty}$ controller for the NDAE state-space model \eqref{NDAE}, we define the following problem statement:
\begin{mypbm} \label{Prob1}
Given the NDAE state-space model \eqref{NDAE} and its extracted LDAE form \eqref{DAE} along with the performance output vector \eqref{zC0} and an imposed controller sparsity structure \eqref{SS}, design an SSOF $\mathcal{H}_{\infty}$ controller.
\end{mypbm}

\subsection{\textcolor{black}{Lyapunov-based approach}} \label{AR}

\textcolor{black}{This section briefly details the Lyapunov-based approach utilized by \cite{nadeem2023robust,nadeem2022robust} to design dense $\mathcal{H}_{\infty}$ controllers and estimators for DAE-modeled power systems.}

\subsubsection{Dense H-infinity controller design}

Considering \eqref{zC0} and LDAE state-space model \eqref{DAE} with $C_y = I_{n_x}, D_y = 0$, dense static state feedback (DSSF) $\mathcal{H}_{\infty}$ controller is \textcolor{black}{proposed in \cite{nadeem2023robust}} for which a convex SDP is cast and solved (as shown in \ref{ApendxA}). To derive the LMIs of the cast convex SDP therein, the $\mathcal{H}_{\infty}$ inequality \eqref{LyaFun} (based on KYP Lemma) is utilized as a cornerstone
 \begin{align}
 \dot{V}(x) + z^T z - \lambda \tilde{w}^T \tilde{w} < 0, \label{LyaFun}
 \end{align}
where $V(x) = x^T E^TP x$ denotes a quadratic Lyapunov candidate function for which $E^TP = P^T E \succeq 0$ holds. Note that to implement the LMIs with strict definiteness, one has to utilize $\epsilon I$-shifting and appropriately choose/tune $\epsilon$ to get a well-performing controller design which is a time-consuming process. 

\subsubsection{Dense H-infinity estimator design}

Considering the following NDAE state-space error dynamics model \cite{nadeem2022robust}:
\begin{align} \label{ErrorD}
    E \dot{e} &= (A+LC_y)e + B_w w + \Delta f,
\end{align}
where $e := \Delta x = x - \hat{x}$ represents the error between the estimated $\hat{x}$ and the true $x$ values of state variables, $L$ denotes the estimator, and $\|\Delta f\|$ is assumed to be Lipschitz bounded by $\alpha \|\Delta x\|$ ($\alpha$: the Lipschitz constant), a dense $\mathcal{H}_{\infty}$ estimator is \textcolor{black}{proposed in \cite{nadeem2022robust}} for which a convex SDP is cast and solved (as shown in \ref{ApendxB}). To derive the LMIs of the cast convex SDP therein, the $\mathcal{H}_{\infty}$ inequality \eqref{LyaFunn} (similarly, based on KYP Lemma) and the $\mathcal{S}$-procedure lemma \cite{derinkuyu2006s} are utilized as cornerstones
 \begin{align} 
 \dot{V}(e) + e^T e - \lambda w^T w < 0, \label{LyaFunn}
 \end{align}
where $V(e) = e^T E^TP e$ denotes a quadratic Lyapunov candidate function for which $E^TP = P^T E \succeq 0$ holds. Likewise, the fine-tuning of $\epsilon I$-shifting in LMIs is inevitable. 

In the next section, we present the proposed methodology to design dense and structured $\mathcal{H}_\infty$ controllers and estimators that ensure the power system is stable in the sense of $\mathcal{H}_\infty$.

\section{\textcolor{black}{Structured H-infinity Designs for DAE-Modeled Power Systems Via NLA}} \label{sec:Met} 

\textcolor{black}{In Section \ref{III-A}, we limit our attention to the derivation of preliminary casting for structured $\mathcal{H}_{\infty}$ controllers for DAE-modeled power systems. Similar derivation for the structured $\mathcal{H}_{\infty}$ estimator design can be obtained as presented in Section \ref{III-C}. Section \ref{III-B} is allocated to elaborate on the utilized non-Lyapunov approach (NLA) and the thorough details of its implementation for the controller design. In Section \ref{III-C}, NLA is similarly implemented to obtain the estimator design.} Procedures \ref{alg:one} and \ref{alg:two} in Sections \ref{III-B} and \ref{III-C}, contain the systematic NLA of designing structured $\mathcal{H}_{\infty}$ controllers and estimators specialized for the DAE-modeled power systems, respectively. Note that as a notation rule, the subscripts $d$ and $a$ refer to the corresponding dynamic and algebraic components, respectively. Also, for the sake of simplicity in derivations, without loss of generality, we assume that $E = \begin{bmatrix}
    I_{n_d} & 0\\0 & 0
\end{bmatrix}$ holds.

\subsection{Preliminary casting for the controller design} \label{III-A}

Let us consider
\begin{align*}
    & A = \begin{bmatrix} A_{dd} & A_{da}\\A_{ad} & A_{aa} \end{bmatrix},~
    B = \begin{bmatrix} B_{d} \\ B_{a} \end{bmatrix},~ 
    C_y = \begin{bmatrix} C_{yd} & C_{ya} \end{bmatrix},\\
    &\mathcal{A}_{ij}(F) := A_{ij} + B_i F C_{yj},~ i,j \in \{d,a\},\\&B_w = \begin{bmatrix} B_{wd}^T & B_{wa}^T \end{bmatrix}^T,~ h = \begin{bmatrix} h_d^T & h_a^T \end{bmatrix}^T,\\
    & B_w^i(F) := B_{wi} + B_i F D_y,~ i \in \{d,a\}.
\end{align*}

According to the algebraic equations of the NDAE state-space model \eqref{NDAE}, we have
\begin{align} \label{xaxd}
    \mathcal{A}_{aa}(F)x_a =&-(\mathcal{A}_{ad}(F) x_d + B_w^a(F) w + h_a).
\end{align}

Assuming the invertibility of $\mathcal{A}_{aa}(F)$ (less restrictive than the invertibility of $A_{aa}$ assumed in \cite{nadeem2023robust}) and utilizing \eqref{xaxd}, we can eliminate $x_a$ from the NDAE state-space model \eqref{NDAE} and get the following ENODE state-space model of the NDAE state-space model \eqref{NDAE}:
\begin{subequations} \label{CNODE}
\begin{align}
& \dot{x}_d = \tilde{A}(F) x_d + \bar{B}_w(F) w + \tilde{h}(F),\\
& \tilde{A}(F) := \mathcal{A}_{dd}(F) - \mathcal{A}_{da}(F) \mathcal{A}_{aa}(F)^{-1} \mathcal{A}_{ad}(F),\\
& \bar{B}_w(F) := B_w^d(F) -\mathcal{A}_{da}(F) \mathcal{A}_{aa}(F)^{-1} B_w^a(F),\\
&\tilde{h}(F) := h_d - \mathcal{A}_{da}(F)\mathcal{A}_{aa}(F)^{-1} h_a.
\end{align}
\end{subequations}

Assuming the $\mathcal{L}_2$-norm boundedness of $h(x,u,w)$ and considering $h(x,u,w)$ as $B_h w_h$, the ENODE state-space model \eqref{CNODE} boils down to the following linear ODE (LODE) form:
\begin{subequations} \label{CODE}
\begin{align}
&\boxed{\dot{x}_d = \tilde{A}(F)x_d + \tilde{B}_w(F) \tilde{w},}\\ 
&\tilde{B}_w(F) := \begin{bmatrix} \bar{B}_w(F) & \bar{B}_h(F) \end{bmatrix},~\bar{B}_h(F) := B_{hd} - \mathcal{A}_{da}(F) \mathcal{A}_{aa}(F)^{-1} B_{ha},\\
& B_h = \begin{bmatrix} B_{hd}^T & B_{ha}^T \end{bmatrix}^T,
\end{align}
\end{subequations}and defining  
\begin{align*}
    & \check{C}(F) := C + D F C_y,~\check{D}_w(F) := D_w + D F D_y,
\end{align*}
the performance output vector \eqref{zC0} boils down to
\begin{subequations} \label{zC}
\begin{align}
& \boxed{z = \tilde{C}(F) x_d + \tilde{D}_w(F) \tilde{w},}\\
& \tilde{C}(F) := \check{C}(F) \begin{bmatrix} I_{n_d}\\ -\mathcal{A}_{aa}(F)^{-1}\mathcal{A}_{ad}(F) \end{bmatrix},~\tilde{D}_w(F) := \begin{bmatrix} \bar{D}_w(F) & \bar{D}_h(F) \end{bmatrix},\\
& \bar{D}_w(F) := \check{D}_w(F) + \check{C}(F) \begin{bmatrix} 0\\ -\mathcal{A}_{aa}(F)^{-1} B_w^a(F) \end{bmatrix},\\
& \bar{D}_h(F) := \check{C}(F) \begin{bmatrix} 0\\ -\mathcal{A}_{aa}(F)^{-1} B_{ha} \end{bmatrix}.
\end{align}
\end{subequations}For LODE \eqref{CODE} and the reduced performance output vector \eqref{zC}, we define the transfer function $T_{z\tilde{w}}(s)$ from the disturbance input vector $\tilde{w}$ to the performance output vector $z$ in \eqref{zC} as follows:
\begin{align*}
T_{z\tilde{w}}(s) &:= \tilde{C}(F)(sI_{n_d} - \tilde{A}(F))^{-1}\tilde{B}_w(F) + \tilde{D}_w(F),
\end{align*}
and denote its $\mathcal{H}_{\infty}$ norm by $\|T_{z\tilde{w}}(s)\|_{\mathcal{H}_{\infty}}$ which is defined as follows \cite{zhou1998essentials}:
\begin{align*}
& \|T_{z\tilde{w}}(s)\|_{\mathcal{H}_{\infty}} := \underset{\Re(s) > 0}{\sup}~\bar{\sigma}[T_{z\tilde{w}}(s)]  \overset{\mathrm{for~a~stable}~T_{z\tilde{w}}(s)}{=} \|T_{z\tilde{w}}(s)\|_{\mathcal{L}_{\infty}} = \underset{\omega \in \mathbb{R}}{\sup}~\bar{\sigma}[T_{z\tilde{w}}(j \omega)]. 
\end{align*}
Note that for an unstable $T_{z\tilde{w}}(s)$ (the closed-loop system with a destabilizing controller $F$, i.e., $\mathbf{sa}(\tilde{A}(F)) \nless 0$) $\|T_{z\tilde{w}}(s)\|_{\mathcal{H}_{\infty}} = \infty$ holds. To compute $\|T_{z\tilde{w}}(s)\|_{\mathcal{H}_{\infty}}$, we utilize the following MATLAB built-in functions:
\begin{subequations} \label{sysF}
\begin{align}
& \boxed{\texttt{sys} = \texttt{ss}(\tilde{A}(F),\tilde{B}_w(F),\tilde{C}(F),\tilde{D}_w(F)),} \\
& \|T_{z\tilde{w}}(s)\|_{\mathcal{H}_{\infty}} = \texttt{hinfnorm}(\texttt{sys}).
\end{align}
\end{subequations}

\subsection{NLA and its implementation for the controller design} \label{III-B}

In this section, we propose a solution to Problem \ref{Prob1} on the basis of NLA. Given the NDAE state-space model \eqref{NDAE}, assuming the $\mathcal{L}_2$-norm boundedness of the linearization error vector associated with the nonlinearities, we extract the LODE state-space model \eqref{CODE} along with the reduced performance output vector \eqref{zC} and utilize NLA to design an SSOF $\mathcal{H}_{\infty}$ controller subject to an imposed sparsity structure encoded by \eqref{SS}. 

\textcolor{black}{As NLA, we utilize the MATLAB built-in function \texttt{hinfstruct} which has been developed based on a structured $\mathcal{H}_{\infty}$ synthesis originally proposed by \cite{apkarian2006nonsmooth}. The proposed synthesis in \cite{apkarian2006nonsmooth} is mainly built upon non-convex non-smooth optimization techniques taking advantage of the Clarke sub-differential \cite{clarke1990optimization} computation of the $\mathcal{H}_{\infty}$ norm. Such a structured $\mathcal{H}_{\infty}$ synthesis has found extensive control applications including but not limited to proportional-integral-derivative (PID) control and combination of proportional-integral (PI) and SOF control of distillation column, aircraft autopilot, and helicopter as highlighted by \cite{gahinet2011structured,gahinet2011decentralized,gahinet2012frequency}. Note that \texttt{hinfstruct} computes the $\mathcal{H}_{\infty}$ norm based on efficient algorithms developed by \cite{boyd1989bisection,boyd1990regularity,gahinet1992numerical}.}

Procedure \ref{alg:one} summarizes NLA to design an SSOF $\mathcal{H}_{\infty}$ controller. In Procedure \ref{alg:one}, $F_0$ denotes the initialization for $F$ and the MATLAB built-in function \texttt{realp}$(F,\textrm{`}F_0\textrm{'})$ creates a real-valued free parameter initialized by $F_0$. To encode the imposed sparsity structure $S$ in \eqref{SS}, we utilize \texttt{F.Free(i,j) = false} for all zero elements of $S$. It is remarkable that extra design specifications can be incorporated via \texttt{options} embedded in \texttt{hinfstruct}. For example, the maximum closed-loop natural frequency and the minimum decay rate for closed-loop poles can be incorporated via \texttt{options.MaxFrequency} and \texttt{options.MinDecay} in \texttt{hinfstruct}, respectively. Also, to avoid high-gain controller designs or unwanted fast dynamics, one may set \texttt{options.MaxFrequency} to a finite value.

\setlength{\floatsep}{5pt}{
\begin{algorithm}[!ht]
\caption{\textbf{SSOF $\mathcal{H}_{\infty}$ Controller Design}}\label{alg:one}
\DontPrintSemicolon
\textbf{input}: $A$, $B$, $B_w$, $B_h$, $C_y$, $D_y$, $C$, $D$, $D_w$, $S$.

\textbf{set} $F_0 = 0$.

\textbf{set} $F$ = \texttt{realp}($F$,`$F_0$').

\For{$i = 1:n_u$}{
\For{$j = 1:n_y$}{
\If{$S(i,j) = 0$}{$\texttt{F.Free}(i,j) = \texttt{false}$.
}}}

\textbf{construct} $\tilde{A}(F)$, $\tilde{B}_w(F)$, $\tilde{C}(F)$, $\tilde{D}_w(F)$ in \eqref{CODE} and \eqref{zC}.

\textbf{construct} \texttt{sys} via \eqref{sysF}.

\textbf{utilize} \texttt{hinfstruct(sys)} to obtain $F^{\ast}$.

\textbf{output}: $F^{\ast}$.

\end{algorithm}}

\textcolor{black}{\begin{myrem}[On \texttt{hinfstruct}] \label{R1}
Borrowing from \cite{gahinet2011structured}, we present a high-level description of the structured $\mathcal{H}_{\infty}$ solver implemented as a building block of \texttt{hinfstruct}. The detailed explanation of the theoretical aspects of \texttt{hinfstruct} can be found in \cite{gahinet2011structured,apkarian2006nonsmooth}.\\Structured $\mathcal{H}_{\infty}$ synthesis deals with the following semi-infinite, non-convex, and non-smooth optimization problem:
\begin{align} \label{hinfs}
    & \underset{K}{\min}~ \|T_{z\tilde{w}}(s)\|_{\mathcal{H}_{\infty}} \iff \underset{\kappa}{\min}~\underset{\omega \in [0,\infty]}{\max}~\bar{\sigma}[T_{z\tilde{w}}(j\omega)],
\end{align}
where $K$ denotes the structured controller/estimator and $\kappa$ represents all the low-level tunable free parameters. It is noteworthy that the function in the right-hand-side of \eqref{hinfs} is the composition of the convex non-smooth function $\max_\omega \circ~ \bar{\sigma}(.)$ with the non-convex differentiable mapping $\kappa \longrightarrow T_{z\tilde{w}}(j\omega)$. Fortunately, such composite functions are Clarke regular \cite{clarke1990optimization} meaning that a complete description of the Clarke subdifferential is available. For the sake of simplicity in representation, \eqref{hinfs} can be rewritten as
\begin{align} \label{fkop}
    & \underset{\kappa}{\min}~[f_{\infty}(\kappa) := \underset{\omega \in [0,\infty]}{\max}~f(\omega,\kappa)].
\end{align}
Clarke regularity ensures that critical point $\kappa^{\ast}$ (usually local minima) are characterized via $0 \in \partial f(\kappa)$. \\To solve \eqref{fkop}, the developers of \texttt{hinfstruct} \cite{gahinet2011structured,apkarian2006nonsmooth} construct a tangent model around the current iterate $\kappa$ that constitutes \textit{quadratic first-order} local approximation of the original problem. An adequate descent direction $\eta$ is then computed by solving the following convex quadratic optimization problem:
\begin{subequations} \label{tangM}
\begin{align}
\underset{\eta}{\min}~&\hat{f}_{\infty}(\kappa + \eta),\\
    \hat{f}_{\infty}(\kappa + \eta) := [\underset{\omega \in \Omega_f}{\max}~&f(\omega,\kappa)] - f_{\infty}(\kappa) + \phi_{\omega}^T \eta + \frac{1}{2} \eta^T Q \eta,
\end{align}
\end{subequations}where $\Omega_f$ denotes some finite set of frequencies, and $\phi_{\omega} \in \partial f(\omega,\kappa)$ represents a subgradient of $f(\omega,\kappa)$. A minimal requirement to implement this scheme is that $\Omega_f$ contains the active frequencies $\omega_{\mathrm{active}}$ achieving the peak value in \eqref{fkop}, namely, $f_{\infty}(\kappa) = f(\omega_{\mathrm{active}},\kappa)$. Such a simple requirement is sufficient for the convergence of the algorithm. Nevertheless, by adding a few extra well-chosen frequencies, one can often improve the quality of the tangent model \eqref{tangM} and take longer steps at each iteration.
\\It is noteworthy that the algorithm implemented in \texttt{hinfstruct} \cite{apkarian2006nonsmooth}, starting from an initial guess $K_{\mathrm{ig}}$ searches for a stabilizing starting point $K_0$ via minimizing the $a$-shifted $\mathcal{H}_{\infty}$ norm \cite{boyd1991linear} of the closed-loop control system where the shift $a > 0$ is usually kept fixed at the initial $a_0 > 0$ for which the $a$-shifted $\mathcal{H}_{\infty}$ norm of the closed-loop control system with $K_{\mathrm{ig}}$ is finite. The authors in \cite{apkarian2006nonsmooth} highlight that the reason for not utilizing the spectral abscissa $\mathbf{sa}$ minimization (to obtain $K_0$) is that minimizing the $a$-shifted $\mathcal{H}_{\infty}$ norm is relatively more compatible with their proposed structured $\mathcal{H}_{\infty}$ synthesis.
\end{myrem}}

\subsection{Casting and NLA implementation for the estimator design} \label{III-C}
To design structured $\mathcal{H}_{\infty}$ estimators for DAE-modeled power systems, a problem similar to as stated in Problem \ref{Prob1} can accordingly be stated. NLA to propose a structured $\mathcal{H}_{\infty}$ estimator is quite the same as the one proposed for an SSOF $\mathcal{H}_{\infty}$ controller. In the sequel, we present a brief set of similar key formulas. To that end, let us consider the following Luenberger-type estimator dynamics:
\begin{subequations} \label{eq:obsr_dynamics}
	\begin{align} 
		\begin{split}
			E\dot{\hat{x}} &= { A}{\hat{x}} +  L\left(  y - \hat{y}\right) + {{B}} {{{u}} } + B_w w_0 + h(\hat{x},u,w),
		\end{split} \label{eq:obsr_dynamics1}\\
		\begin{split}
			\hat{ y} &= C \hat{x},
		\end{split} \label{eq:obsr_dynamics2}
	\end{align}
\end{subequations}
where $\hat{x}$ and $\hat{y}$ denote the estimated states and outputs, respectively and $w_0$ represents the steady-state value of $w$ which contains load demands and renewable generations. Now, defining the error as $e := x - \hat{x}$, then, the model of error dynamics can be computed as given in \eqref{ErrorD}. The main idea is designing $L$ such that the error model \eqref{ErrorD} asymptotically converges to zero. To that end, considering
\begin{align*}
    & e = \begin{bmatrix} e_d^T & e_a^T \end{bmatrix}^T,~ \mathcal{A}_{ij}(L) := A_{ij} + L_i C_{yj},~ i,j \in \{d,a\},\\
 & \Delta f = \begin{bmatrix} {\Delta f}_d^T & {\Delta f}_a^T \end{bmatrix}^T,~L = \begin{bmatrix} L_d^T & L_a^T \end{bmatrix}^T,~C_y = \begin{bmatrix} C_{yd} & C_{ya} \end{bmatrix},
\end{align*}
and assuming the invertibility of $\mathcal{A}_{aa}(L)$ to eliminate $e_a$ from the NDAE state-space model \eqref{ErrorD} via  
\begin{align*}
\mathcal{A}_{aa}(L) e_a &= -(\mathcal{A}_{ad}(L)e_d + B_{wa} w + {\Delta f}_a), 
\end{align*}
we get the following ENODE state-space model of the NDAE state-space model \eqref{ErrorD}:
\begin{subequations} \label{EDO}
\begin{align} 
    & \dot{e}_d = \tilde{A}(L) e_d + \bar{B}_w(L) w + \Delta \tilde{f}(L),\\
& \tilde{A}(L) := \mathcal{A}_{dd}(L) - \mathcal{A}_{da}(L) \mathcal{A}_{aa}(L)^{-1} \mathcal{A}_{ad}(L),\\
 & \bar{B}_w(L) := B_{wd} - \mathcal{A}_{da}(L) \mathcal{A}_{aa}(L)^{-1} B_{wa},\\
 & \Delta \tilde{f}(L) := {\Delta f}_d - \mathcal{A}_{da}(L)\mathcal{A}_{aa}(L)^{-1} {\Delta f}_a.
\end{align}  
\end{subequations}

Assuming the $\mathcal{L}_2$-norm boundedness of $\Delta f$ and considering it as $B_{\Delta f} w_{\Delta f}$, the ENODE state-space model \eqref{EDO} reduces to the following LODE form:
\begin{subequations} \label{LEDO}
\begin{align} 
    & \boxed{\dot{e}_d = \tilde{A}(L) e_d + \tilde{B}_w(L) \tilde{w},}\\
    &\tilde{B}_w(L) := \begin{bmatrix} \bar{B}_w(L) & \bar{B}_{\Delta f}(L) \end{bmatrix},\\& 
\bar{B}_{\Delta f}(L) := B_{\Delta f d} - \mathcal{A}_{da}(L) \mathcal{A}_{aa}(L)^{-1} B_{\Delta f a},~B_{\Delta f} = \begin{bmatrix} B_{\Delta f d}^T & B_{\Delta f a}^T \end{bmatrix}^T,
\end{align} 
\end{subequations}and the performance output vector $z = e$ reduces to
\begin{subequations} \label{ZE}
\begin{align} 
    & \boxed{z = \tilde{C}(L) e_d + \tilde{D}_w(L)\tilde{w},}\\
& \tilde{C}(L) := \begin{bmatrix} I_{n_d}\\ -\mathcal{A}_{aa}(L)^{-1}\mathcal{A}_{ad}(L) \end{bmatrix},~\tilde{D}_w(L) := \begin{bmatrix} \bar{D}_w(L) & \bar{D}_{\Delta f}(L) \end{bmatrix},\\
& \bar{D}_w(L) := \begin{bmatrix} 0\\ -\mathcal{A}_{aa}(L)^{-1} B_{wa} \end{bmatrix},~\bar{D}_{\Delta f}(L) := \begin{bmatrix} 0\\ -\mathcal{A}_{aa}(L)^{-1} B_{\Delta f a} \end{bmatrix}.
\end{align}
\end{subequations}For LODE \eqref{LEDO} and the reduced performance output vector \eqref{ZE}, we have
\begin{subequations} \label{sysL}
\begin{align}
& \boxed{\texttt{sys} = \texttt{ss}(\tilde{A}(L),\tilde{B}_w(L),\tilde{C}(L),\tilde{D}_w(L)),}\\
& \|T_{z\tilde{w}}(s)\|_{\mathcal{H}_{\infty}} = \texttt{hinfnorm}(\texttt{sys}). \notag
\end{align}    
\end{subequations}Note that we can impose a sparsity structure to the estimator $L$ via $L = L \odot S$ similar to as in \eqref{SS}. With a slight modification to Procedure \ref{alg:one}, we obtain a similar Procedure to design a structured $\mathcal{H}_{\infty}$ estimator, \textcolor{black}{namely,} Procedure \ref{alg:two}, where $L_0$ denotes the initialization for $L$.

\setlength{\floatsep}{5pt}{\begin{algorithm}[!ht]
\caption{\textbf{Structured $\mathcal{H}_{\infty}$ Estimator Design}}\label{alg:two}

\textbf{input}: $A$, $C_y$, $B_w$, $B_{\Delta f}$, $S$.

\textbf{set} $L_0 = 0$.

\textbf{set} $L$ = \texttt{realp}($L$,`$L_0$').

\For{$i = 1:n_x$}{
\For{$j = 1:n_y$}{
\If{$S(i,j) = 0$}{$\texttt{L.Free}(i,j) = \texttt{false}$.}}}

\textbf{construct} $\tilde{A}(L)$, $\tilde{B}_w(L)$, $\tilde{C}(L)$, $\tilde{D}_w(L)$ in \eqref{CODE} and \eqref{zC}.

\textbf{construct} \texttt{sys} via \eqref{sysL}.

\textbf{utilize} \texttt{hinfstruct(sys)} to obtain $L^{\ast}$.

\textbf{output}: $L^{\ast}$.

\end{algorithm}}

\section{Numerical Simulations}\label{sec:Example}

In this section, we assess the effectiveness of NLA in designing structured controllers and estimators for the DAE-modeled power systems. We design various types of controllers and estimators with different sparsity structures such as centralized (dense), decentralized, and distributed architectures. Notice that in the proposed methodology, this can easily be achieved by appropriately selecting the sparsity structure $S$. With that in mind, we assess the performance of the proposed methodology on four IEEE test systems with the following specifications: \textit{(i)} IEEE $9$-bus with $(n_x,n_d,n_a,n_u,n_w) = (36,12,24,6,18)$, \textit{(ii)} IEEE $14$-bus with $(n_x,n_d,n_a,n_u,n_w) = (58,20,38,10,28)$, \textit{(iii)} IEEE $39$-bus with $(n_x,n_d,n_a,n_u,n_w) = (138,40,98,20,78)$, and \textit{(iv)} IEEE $57$-bus with $(n_x,n_d,n_a,n_u,n_w) = (156,28,128,14,114)$. To compute the proposed structured $\mathcal{H}_{\infty}$ controllers and estimators, we run Procedure \ref{alg:one} and Procedure \ref{alg:two}, respectively. All the numerical experiments have been run in MATLAB R$2022$b on a MacBook Pro with a $3.1$ GHz Intel Core i$5$ and memory $8$ GB $2133$ MHz. For solving the convex SDPs, we have utilized CVX \cite{grant2014cvx} to implement the LMIs in MOSEK \cite{andersen2000mosek} SDP solver. We simply set $D_w = 0$ for all the numerical experiments.

Through the comprehensive case studies in this section, we seek answers to the following questions:
\begin{itemize}
    \item \textcolor{black}{\textit{Q1}: How efficient is NLA in comparison with the Lyapunov-based approach in terms of computation scalability and computation sensitivity?}    
    
    \item \textcolor{black}{\textit{Q2}: Under what conditions NLA can potentially face a scalability issue leading to even underperformance compared to the Lyapunov-based approach?}
    
    \item \textcolor{black}{\textit{Q3}: Can NLA be utilized to propose $\mathcal{H}_{\infty}$ controller designs subject to the imposed structural/sparsity constraints with partially-accessed states in the presence of noisy measurement outputs?}
    
    \item \textit{Q4}: How can output matrix selection affect the $\mathcal{H}_{\infty}$ performance and the computational time corresponding to the SSOF controller design built upon NLA?
    
    \item \textcolor{black}{\textit{Q5}: What advantages can be achieved via the proposed structured $\mathcal{H}_{\infty}$ estimator design built upon NLA?}
\end{itemize}

\begin{figure}[ht]
\centering

\subfloat{\includegraphics[keepaspectratio=true,scale=0.7]{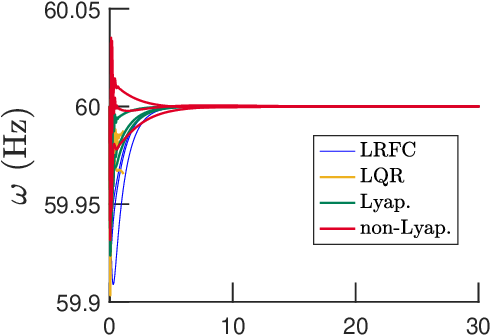}}{}{}
\subfloat{\includegraphics[keepaspectratio=true,scale=0.7]{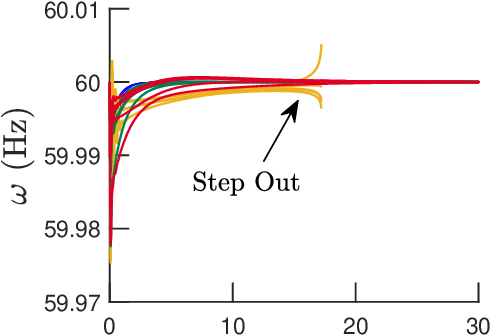}}{}{}
\\
\subfloat{\includegraphics[keepaspectratio=true,scale=0.7]{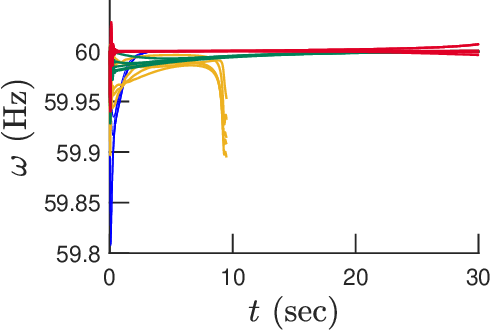}}{}{}
\subfloat{\includegraphics[keepaspectratio=true,scale=0.7]{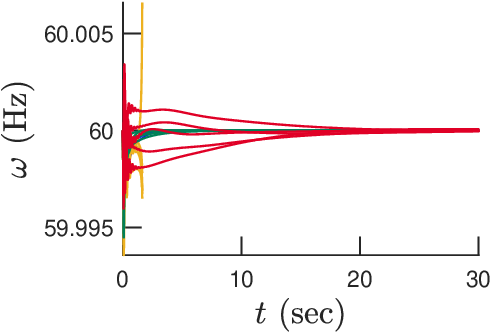}}{}{}

\caption{\textcolor{black}{(Section \ref{Sec:A}) The generator frequencies for $9$-bus (top-left), $14$-bus (top-right), $39$-bus (bottom-left), and $57$-bus (bottom-right) test systems, for moderate disturbance in load demand and renewable power generation.}}\label{fig:case_low}
\end{figure}

\begin{figure}[ht]
\centering

\subfloat{\includegraphics[keepaspectratio=true,scale=0.7]{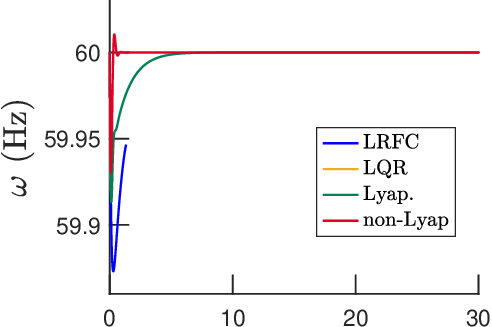}}{}{}
\subfloat{\includegraphics[keepaspectratio=true,scale=0.7]{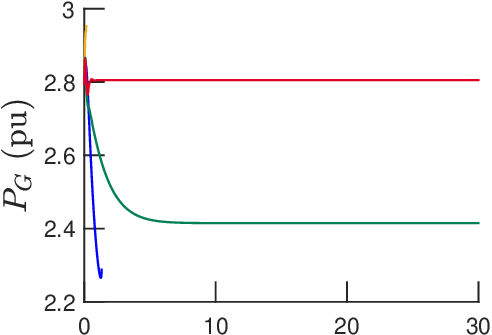}}{}{}
\\
\subfloat{\includegraphics[keepaspectratio=true,scale=0.7]{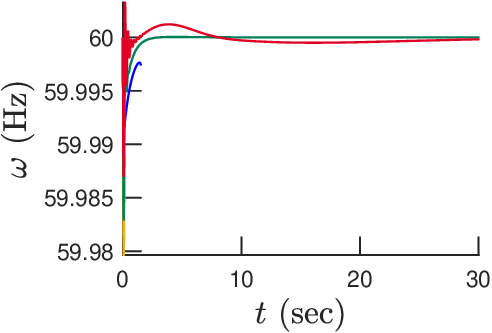}}{}{}
\subfloat{\includegraphics[keepaspectratio=true,scale=0.7]{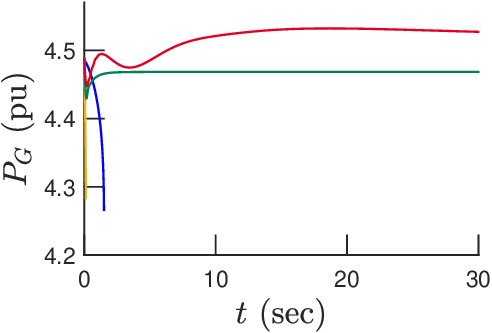}}{}{}

\caption{\textcolor{black}{(Section \ref{Sec:A}) Generator $3$ frequency and power for $9$-bus (top) and $57$-bus (bottom) test systems, under large disturbance in load demand and renewable power generation.}}\label{fig:case_high}
\end{figure}

\subsection{\textcolor{black}{Dense controller design with fully-accessed states and noiseless measurement states}} \label{Sec:A}

In this section, we design a full-state DSSF $\mathcal{H}_{\infty}$ controller with no noise in the measurement states. This can be achieved by setting $S = \mathbf{1}_{n_u \times n_y}$, $C_y = I_{n_x}$, and $D_y = 0$. Furthermore, we consider $B_h = 0.1I_{n_x}$.
\textcolor{black}{To showcase the superiority of the proposed methodology, we compare the performance of the proposed methodology with the conventional LQR-type controller presented in \cite{sadamoto2019dynamic} and the Lyapunov-based controllers given in \cite{nadeem2023robust,nugroho2023load} for the various IEEE test systems.} 

\textcolor{black}{We observe from Tab. \ref{table:LNL} that as the size of the system increases, the computational time for calculating the controller matrix $F$ using the Lyapunov-based approach \cite{nadeem2023robust} increases significantly while for NLA, the computational time is much less and is more than $47$ ($\frac{5361.67-111.31}{111.31} \approx 47.17$) times faster on the largest IEEE test system, i.e., IEEE $57$-bus. We can also see that the proposed methodology yields exactly the same $\mathcal{H}_\infty$ performance value as compared to \cite{nadeem2023robust}. This means that the performance of the proposed controller is not degraded and hence provides the same stability improvement on power systems after disturbances.}  

\textcolor{black}{Also, we observe that unlike the Lyapunov-based approach \cite{nadeem2023robust}, NLA controller is not a high-gain controller which makes it more suitable for practical purposes. It is noteworthy that the corresponding values of the Lyapunov-based approach \cite{nadeem2023robust} reported in Tab. \ref{table:LNL} are obtained after fine-tuning of $\epsilon I$-shifting of the corresponding LMIs (setting $\epsilon = 10^{-3.3}$) which is a time-consuming process by itself. Nevertheless, we have not taken into account the time of fine-tuning in computational times reflected in Tab. \ref{table:LNL}. Moreover, through the numerical experiments, we observe that although the approach proposed in \cite{nadeem2023robust} is capable of designing the dense $\mathcal{H}_{\infty}$ controllers with the same $\mathcal{H}_{\infty}$ performance level as the presented NLA, there is a noticeable gap between the value of $\lambda := \mu^2$ in the corresponding convex SDP and the true $\mathcal{H}_{\infty}$-squared value i.e., $\|T_{z\tilde{w}}(s)\|_{\mathcal{H}_{\infty}}^2$ evaluated by \texttt{hinfnorm}. Therefore, we have reported the corresponding true $\mathcal{H}_{\infty}$ values in Tab. \ref{table:LNL}.} 

That being said, we now assess the performance of NLA controller in stabilizing the power system under transient conditions. To that end, large disturbances in load and renewable (which are modeled as negative loads) have been added and the simulations are performed as follows: Initially, the system load demand and power generation are exactly equal, and thus the system rests in equilibrium conditions. Then, immediately after $t>0$, there is an abrupt change in load demand and renewable power generation, and their new values are given as follows:
\begin{align*}
& P'_L + Q'_L = (1+\Delta_L)(P^0_L + Q^0_L),~P'_R + Q'_R =(1+\Delta_R)(P^0_L + Q^0_L),
\end{align*}
where $P^0_L$, $Q^0_L$, $P^0_R$, and $Q^0_R$ denote the initial values of active/reactive load demand and renewable generation, while $P'_L$, $Q'_L$, $P'_R$, and $Q'_R$ denote their values after the disturbances, and
$\Delta_L$ and $\Delta_R$ denote the severity of the disturbances in load and renewable power generation, respectively. We apply different severity of disturbances by choosing different values of $\Delta_L$, and $\Delta_R$. To that end, we run two simulation studies for all the IEEE test systems. For $9$-bus test system, we choose $(\Delta_L,\Delta_R)= (0.5,-0.3)$ for the first case study and $(\Delta_L,\Delta_R)= (0.9,-0.5)$ for the second case study. Similarly, for $14$-bus and $39$-bus test systems, we respectively select $(\Delta_L,\Delta_R)= (0.05,-0.03)$ and $(\Delta_L,\Delta_R)= (0.15,-0.15)$, while for $57$-bus system, we use $(\Delta_L,\Delta_R)= (0.01,-0.01)$ for the first case study and $(\Delta_L,\Delta_R)= (0.05,-0.05)$ for the second case study. 

\textcolor{black}{The results are presented in Figs. \ref{fig:case_low} and \ref{fig:case_high}. Notice that in all the visualizations, we refer to the controller proposed in \cite{nugroho2023load} as LRFC (load and renewable following control) and to the one proposed in \cite{nadeem2023robust} as Lyapunov-based (abbreviated as Lyap) while to the controller proposed in the current paper as NLA (non-Lyap). 
From Fig. \ref{fig:case_low}, we observe that for a moderate disturbance in load and renewable, the LQR-type controller proposed in \cite{sadamoto2019dynamic} is unable to stabilize the system while the Lyapunov-based controllers \cite{nadeem2023robust,nugroho2023load} (note that LRFC is also classified as a Lyapunov-based controller) and NLA controller can successfully keep the system stable and synchronized.} 

\textcolor{black}{Furthermore, from Fig. \ref{fig:case_high}, we observe that as the severity of the disturbances in load and renewable increases, only the controller proposed in \cite{nadeem2023robust} and the presented controller in this paper are able to keep the system synchronized. This corroborates the result presented in Tab. \ref{table:LNL} from which, we see that the Lyapunov-based approach and presented NLA $\mathcal{H}_\infty$ attain almost the same $\mathcal{H}_\infty$ value, thus, providing almost the same performance. However, the main benefit of the presented approach over \cite{nadeem2023robust} is that it is far less computationally cumbersome. This holds because unlike \cite{nadeem2023robust} no convex SDP needs to be cast and solved to compute the controller matrix $F$.} Moreover, the reason behind the fact that controllers in \cite{nugroho2023load,sadamoto2019dynamic} are unable to keep the power system synchronized, is that the disturbance in load and renewable are not modeled in the controller architecture. In both of these studies, the disturbances are assumed to be zero in the controller design thus making them less robust. 

\begin{table}[t]
\centering
\caption{\textcolor{black}{(Section \ref{Sec:A}) The $\mathcal{H}_{\infty}$ values and computational times for the Lyapunov-based approach \cite{nadeem2023robust} and NLA corresponding to DSSF $\mathcal{H}_{\infty}$ controller with $S = \mathbf{1}_{n_u \times n_y}$, $C_y = I_{n_x}$, $D_y = 0$, and $B_h = 0.1I_{n_x}$ for the IEEE test systems.}}
\label{table:LNL}
\begin{tabular}{|c|c|c|c|}
\hline
\textrm{Approach} & $\|T_{z\tilde{w}}(s)\|_{\mathcal{H}_{\infty}}$ & Computational Time & $(n_u,n_y)$ \\
\hline
non-Lyap $9$-bus & \cellcolor{lightgray}$3.1856$ & \cellcolor{lightgray} $2.09$ s & $(6,36)$\\
\hline
Lyap $9$-bus & $3.1857$ & $2.72$ s & $(6,36)$\\
\hline
non-Lyap $14$-bus & \cellcolor{lightgray}$7.4260$ & \cellcolor{lightgray} $6.05$ s & $(10,58)$\\
\hline
Lyap $14$-bus & $7.4274$ & $19.41$ s & $(10,58)$\\
\hline
non-Lyap $39$-bus & $9.2018$ & \cellcolor{lightgray} $259.56$ s & $(20,138)$\\
\hline
Lyap $39$-bus & \cellcolor{lightgray}$9.1701$ & $1611.45$ s & $(20,138)$\\
\hline
non-Lyap $57$-bus & \cellcolor{lightgray}$23.5716$ & \cellcolor{lightgray} $111.31$ s & $(14,156)$\\
\hline
Lyap $57$-bus & $23.5731$ & $5361.67$ s & $(14,156)$\\
\hline
\end{tabular}
\end{table}

\subsection{\textcolor{black}{Structured controller design with partially-accessed states and noisy measurement outputs}} \label{Sec:B}

\textcolor{black}{In the previous section, we designed a full-state feedback controller, meaning that the control action taken by each generator also depends on all the rest of the generators and network states. However, this could be unrealistic and require a dense and reliable communication network. To that end, here, we design an SSOF $\mathcal{H}_\infty$ controller and simultaneously consider the scenario in which, we have noisy measurement outputs. In the proposed methodology, various sparsity structures (such as decentralized and distributed) on the controller matrix $F$ can be imposed using $S$ as discussed in Procedure \ref{alg:one}. Here, we design two types of controller architecture: \textit{(i)} completely decentralized design, meaning that each generator utilizes only its own state information, and \textit{(ii)} distributed design in which the control action taken by each generator depends on a few other state information. Furthermore, to consider noisy output measurements, we set $D_y = 0.1I_{n_y,n_w}$.} 

\textcolor{black}{Tab. \ref{table:NL} represents the $\mathcal{H}_{\infty}$ values and computational times for NLA corresponding to SSOF $\mathcal{H}_{\infty}$ controllers for different sparsity structures for the IEEE test systems. Such different sparsity structures are three-fold as follows:}
\textcolor{black}{\begin{itemize}
    \item \textit{Centralized}: in this case, we set $S$ as the following dense sparsity structure:
    \begin{align*}
        S &= \mathbf{1}_{n_u \times n_y}.
    \end{align*}
    \item \textit{Distributed}: in this case, we set $S$ as the following randomly generated zero-one sparsity structure:
    \begin{align*}
        S &= \texttt{randi}(\begin{bmatrix}
            0 & 1
        \end{bmatrix},n_u,n_y).
    \end{align*}
    where $\texttt{randi}(\begin{bmatrix}
            0 & 1
        \end{bmatrix},n_u,n_y)$ is a MATLAB built-in function that generates a $n_u$ by $n_y$ matrix consisting of uniformly distributed pseudorandom zero-one values.
    \item \textit{Decentralized}: in this case, noting that $(n_u,n_d) = (2 N,4N)$ holds, we choose $C_y = \begin{bmatrix} I_{n_d} & 0 \end{bmatrix}$ and set $S$ as the following decentralized sparsity structure:
    \begin{align*}
        S &= \begin{bmatrix}
            I_{N} & I_{N} & I_{N} & I_{N}\\
            I_{N} & I_{N} & I_{N} & I_{N}
        \end{bmatrix} = \mathbf{1}_{2 \times 4} \otimes I_{N}.
    \end{align*}
\end{itemize}}

\begin{figure}[t]
    \centering
    \includegraphics[trim = {1cm 3cm 0cm 3cm}, clip, scale = 0.35]{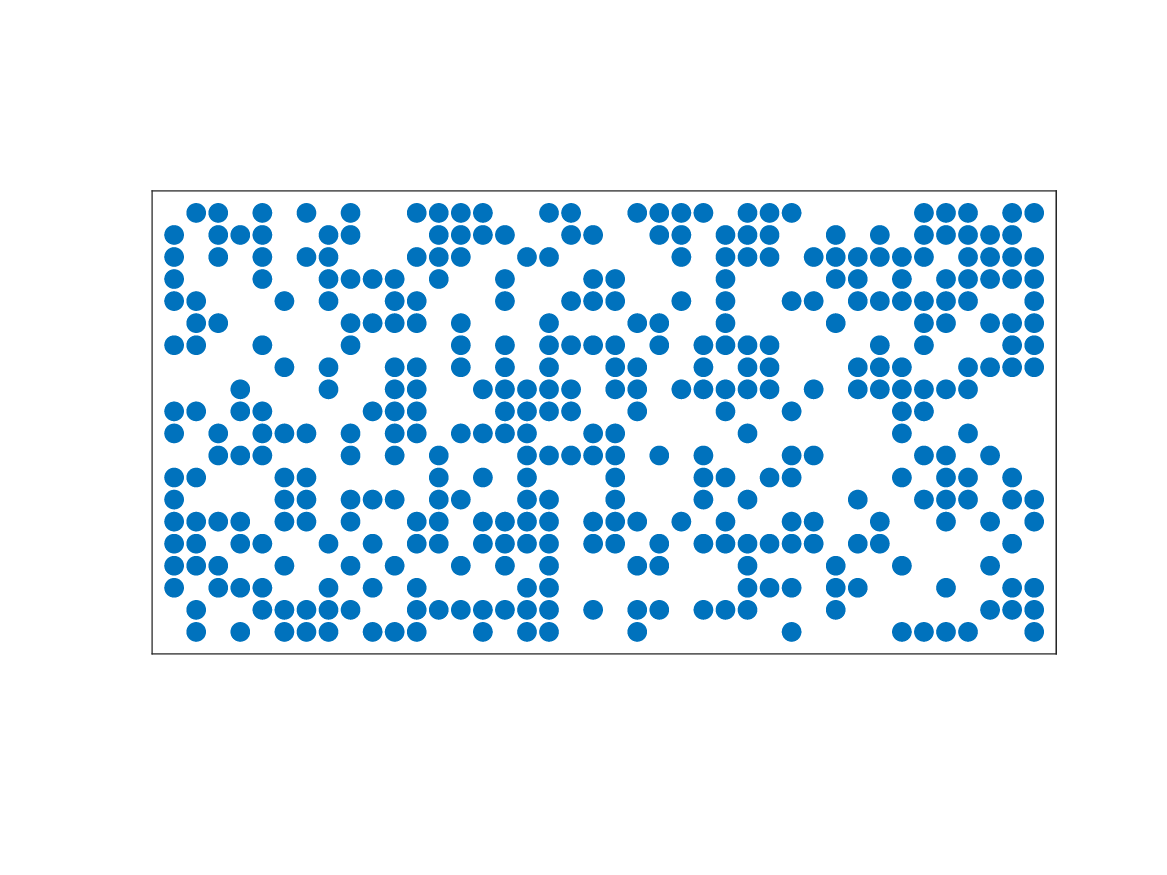}
    \caption{\textcolor{black}{(Section \ref{Sec:B}) The randomly generated zero-one sparsity structure of the Distributed $39$-bus $F$ reported in Tab. \ref{table:NL}. The blue dots represent the nonzero values of $F$.}}
    \label{fig:spp}
\end{figure}

\textcolor{black}{Notice that the smaller the $\mathcal{H}_{\infty}$ value is, the better the controller performance is. We see from Tab. \ref{table:NL} that as sparsity of the controller is promoted the corresponding $\mathcal{H}_{\infty}$ value increases which makes sense because the controller loses some state information. In other words, there exists a fundamental trade-off between the $\mathcal{H}_{\infty}$ performance and the controller sparsity. Fig. \ref{fig:spp} depicts the randomly generated zero-one sparsity structure of the Distributed $39$-bus $F$ reported in Tab. \ref{table:NL}.}

\begin{figure}[ht]
\centering

\subfloat{\includegraphics[keepaspectratio=true,scale=0.7]{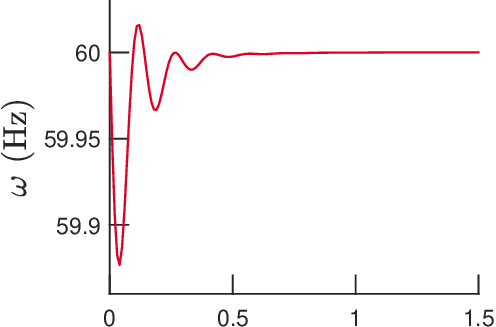}}{}{}
\subfloat{\includegraphics[keepaspectratio=true,scale=0.7]{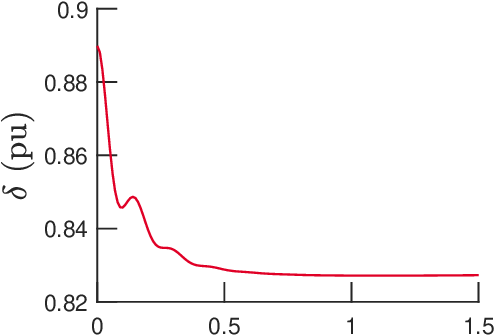}}{}{}
\\
\subfloat{\includegraphics[keepaspectratio=true,scale=0.7]{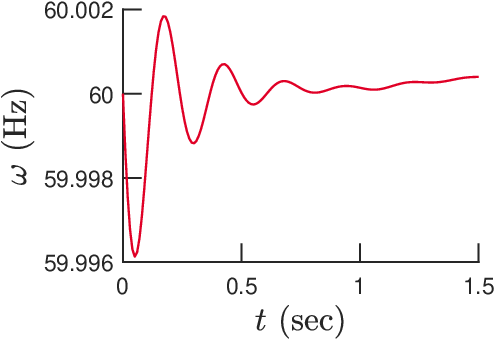}}{}{}
\subfloat{\includegraphics[keepaspectratio=true,scale=0.7]{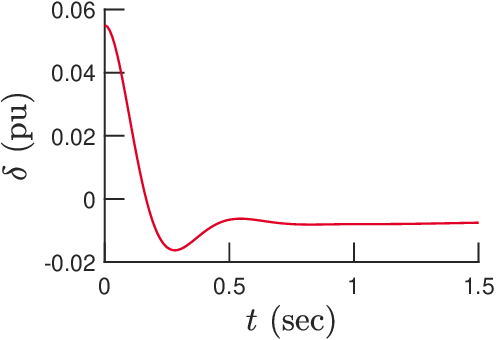}}{}{}

\caption{\textcolor{black}{(Section \ref{Sec:B}) The dynamic performance of the proposed SSOF $\mathcal{H}_{\infty}$ controller under large disturbance in load demand and renewable power generation and noisy measurement states; Generator $1$ frequency and rotor angle for $9$-bus (top) and $57$-bus (bottom) test systems.}}\label{fig:case_high_}
\end{figure}

\textcolor{black}{The dynamic performance of the system is assessed under load and renewable disturbance as discussed in the previous section. Note that the Lyapunov-based approaches proposed in \cite{nadeem2023robust, nugroho2023load,sadamoto2019dynamic} are not capable of designing an SSOF $\mathcal{H}_{\infty}$ controller, thus their results are not included here. The frequencies of all the generators are presented in Fig. \ref{fig:case_high_}. We see that after a large abrupt disturbance in load and renewable, NLA controller can successfully keep the system transiently stable and synchronized. 
We can also see that the non-Lyap controller can also bring the system back to its nominal frequency after the large disturbance.}

\begin{table}[t]
\centering
\caption{\textcolor{black}{(Section \ref{Sec:B}) The $\mathcal{H}_{\infty}$ values and computational times for NLA corresponding to SSOF $\mathcal{H}_{\infty}$ controllers with different choices of $S$: Centralized (Cen.), Distributed (Dis.), and Decentralized (Dec.), $C_y = I_{n_y,n_x}$, $D_y = 0.1I_{n_y,n_w}$, $n_y = n_d$, and $B_h = 0.1I_{n_x}$ for the IEEE test systems.}}
\label{table:NL}
\begin{tabular}{|c|c|c|c|}
\hline
\textrm{Sparsity Structure} & $\|T_{z\tilde{w}}(s)\|_{\mathcal{H}_{\infty}}$ & Computational Time & $(n_u,n_y)$ \\
\hline
Cen. $9$-bus & \cellcolor{lightgray}$3.2314$ & \cellcolor{lightgray}$1.69$ s & $(6,12)$\\
\hline
Dis. $9$-bus & $3.3007$ & $4.86$ s & $(6,12)$\\
\hline
Dec. $9$-bus & $3.2784$ & $4.11$ s & $(6,12)$\\
\hline
Cen. $14$-bus & \cellcolor{lightgray}$7.3964$ & $2.04$ s & $(10,20)$\\
\hline
Dis. $14$-bus & $7.4446$ & $5.40$ s & $(10,20)$\\
\hline
Dec. $14$-bus & $7.4274$ & \cellcolor{lightgray}$1.78$ s & $(10,20)$\\
\hline
Cen. $39$-bus & \cellcolor{lightgray}$9.2120$ & \cellcolor{lightgray}$42.59$ s & $(20,40)$\\
\hline
Dis. $39$-bus & $9.9968$ & $145.53$ s & $(20,40)$\\
\hline
Dec. $39$-bus & $9.8467$ & $82.38$ s & $(20,40)$\\
\hline
Cen. $57$-bus & \cellcolor{lightgray}$23.6002$ & \cellcolor{lightgray}$35.33$ s & $(14,28)$\\
\hline
Dis. $57$-bus & $23.6168$ & $51.04$ s & $(14,28)$\\
\hline
Dec. $57$-bus & $23.6180$ & $56.93$ s & $(14,28)$\\
\hline
\end{tabular}
\end{table}

\begin{myrem}
Note that unlike the \cite{nadeem2023robust}, we take into account a more general form for the nonlinearity $h$ by choosing $B_h \propto I_{n_x}$ instead of $B_h \propto B_w$ as in the former one, the number of columns of $B_h$ is $n_x$ while in the latter one, the number of columns of $B_h$ is $n_w < n_x$. Then, such a more general choice of $B_h$ makes the Lyapunov-based approach even more inefficient in terms of scalability as it makes the LMI in the convex SDP even larger by adding extra rows and columns.
\end{myrem}

\begin{figure}[ht]
\centering

\subfloat{\includegraphics[keepaspectratio=true,scale=0.35]{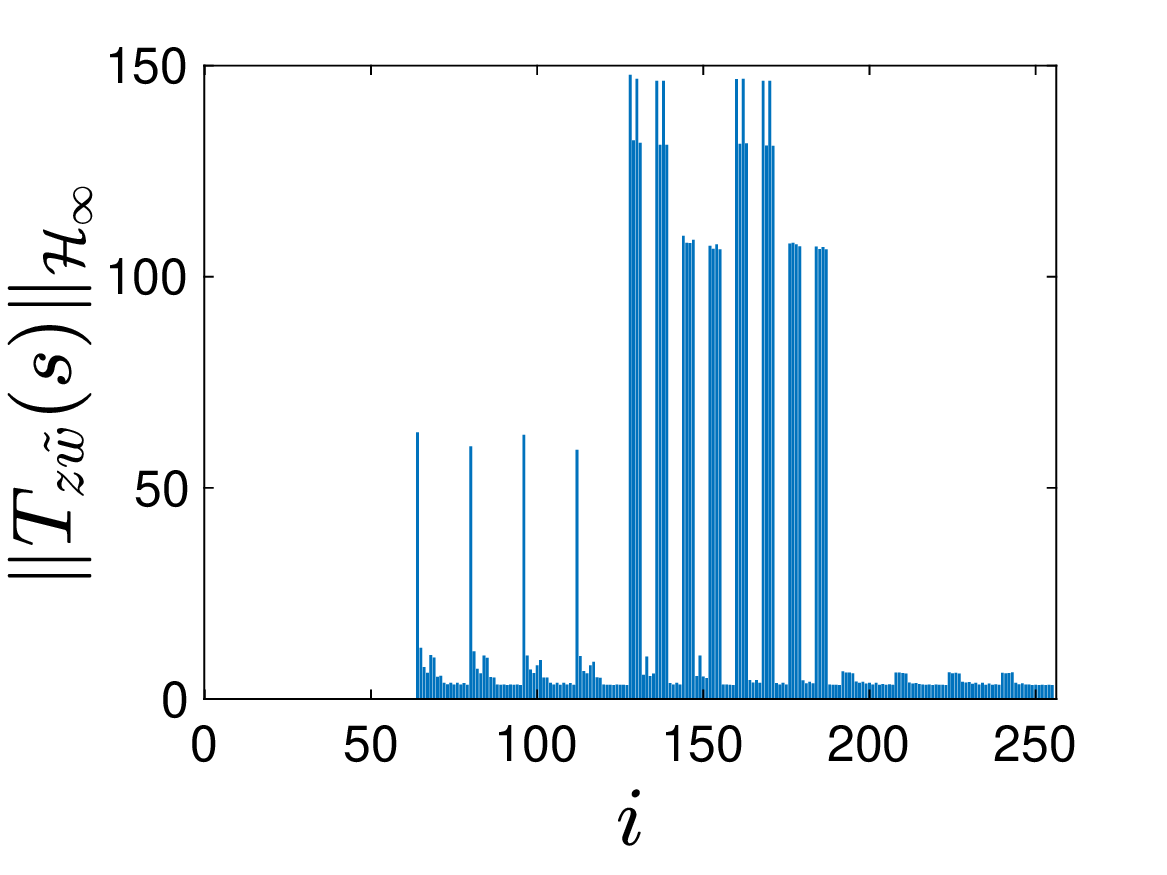}}{}{}
\subfloat{\includegraphics[keepaspectratio=true,scale=0.35]{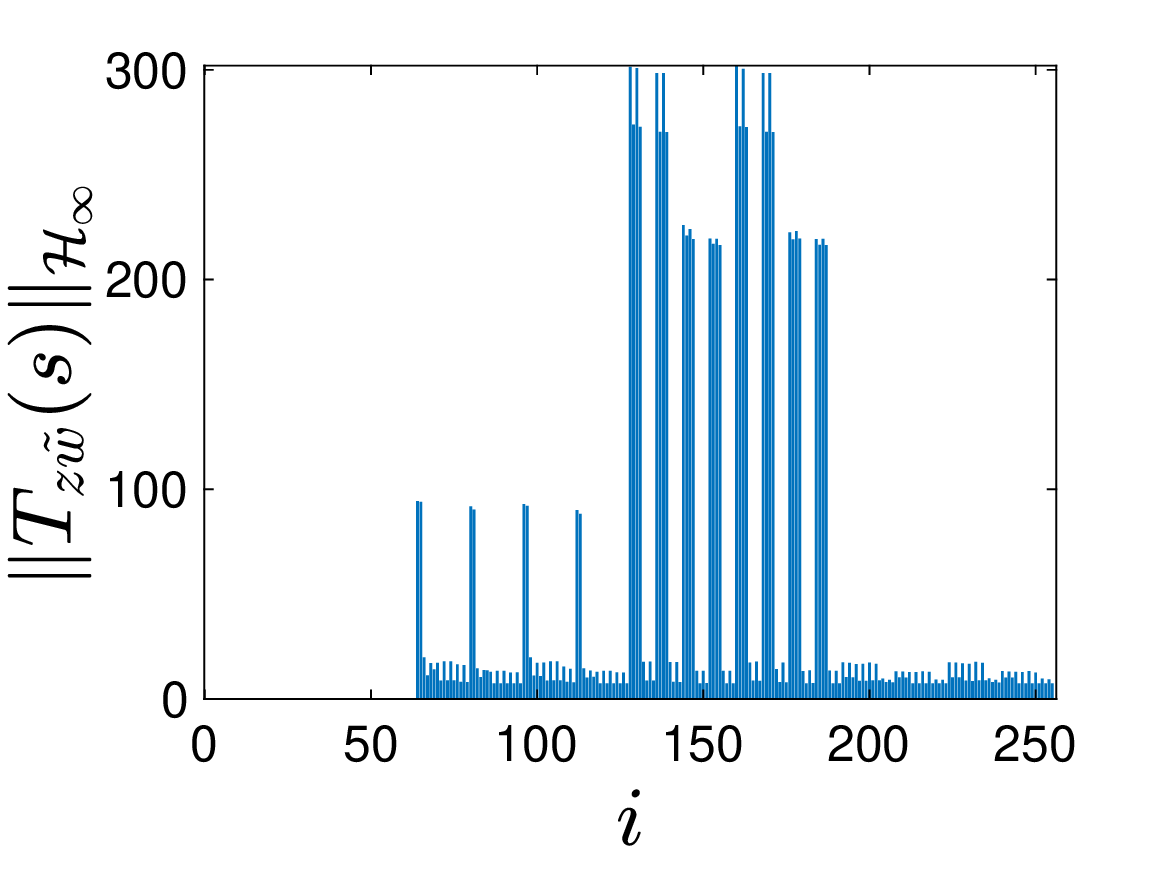}}{}{}
\\
\subfloat{\includegraphics[keepaspectratio=true,scale=0.35]{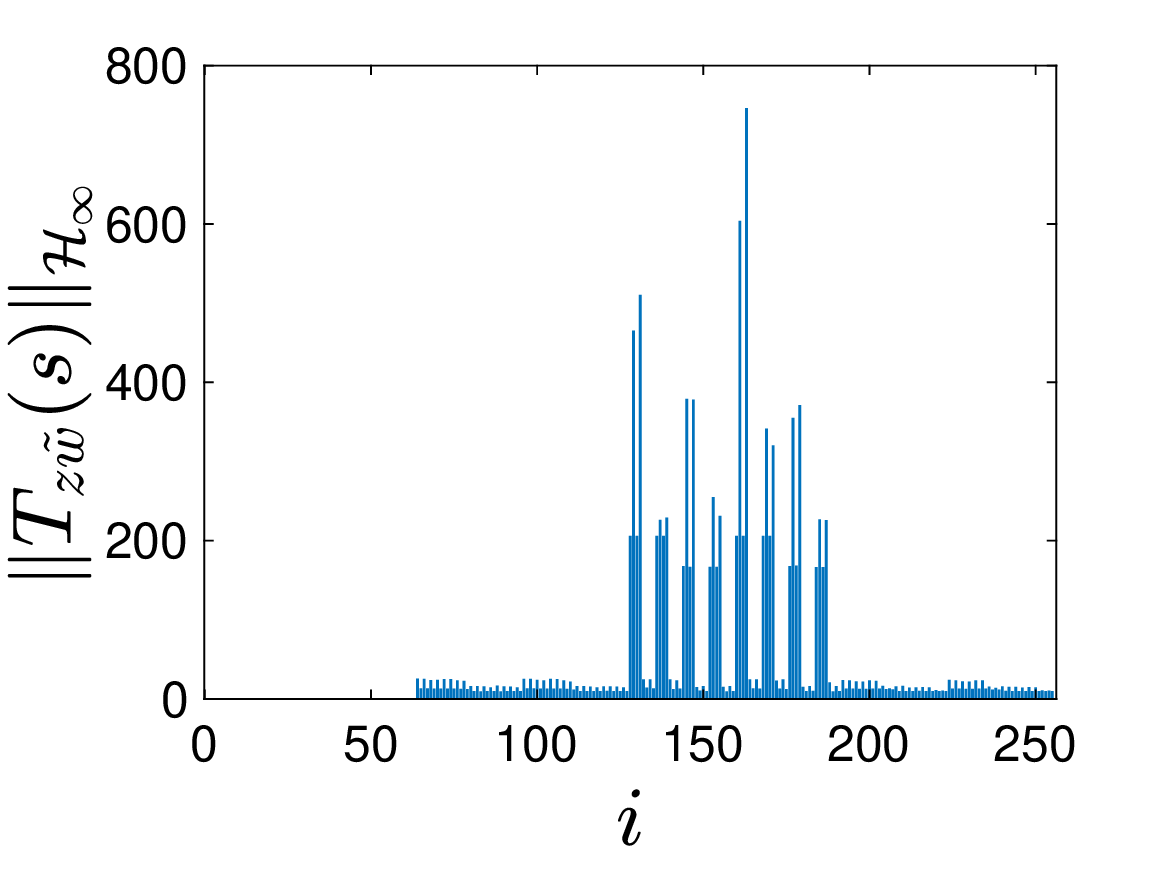}}{}{}
\subfloat{\includegraphics[keepaspectratio=true,scale=0.35]{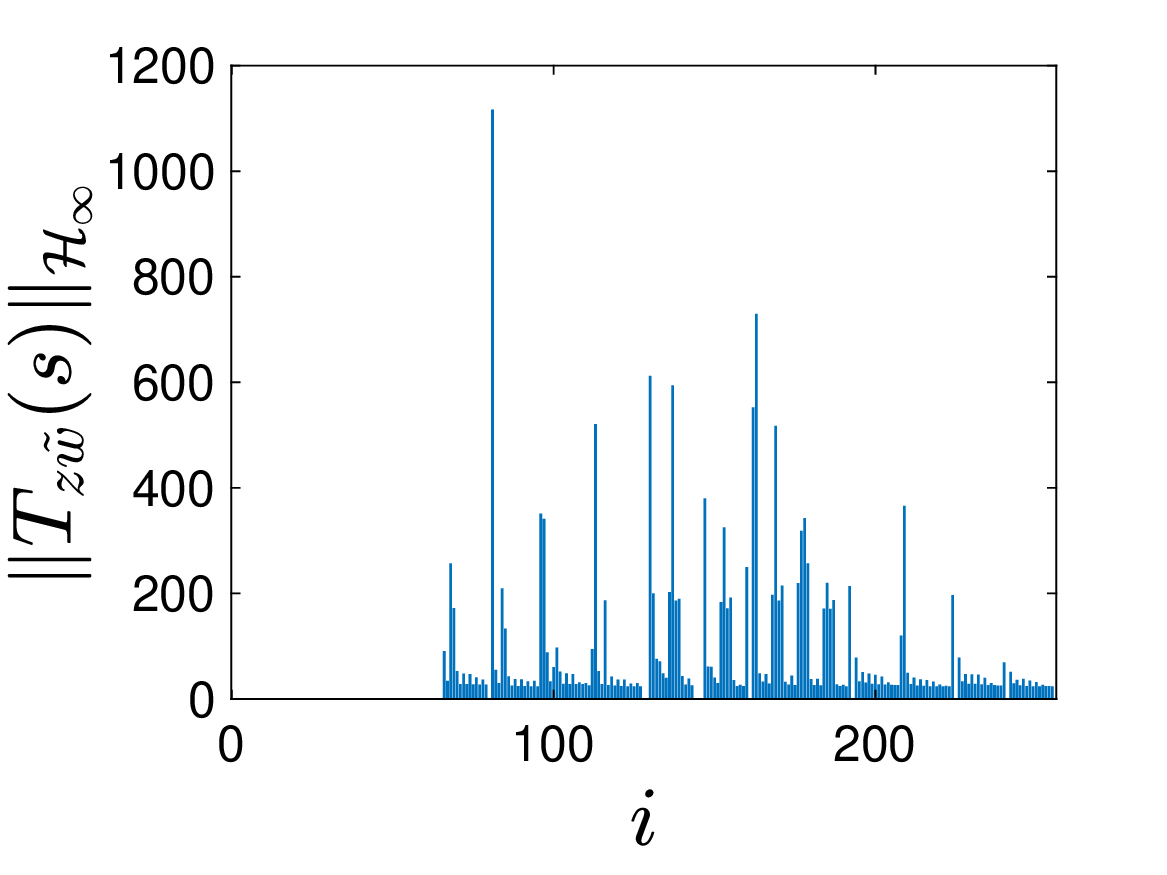}}{}{}

\caption{(Section \ref{Sec:B}) The $\mathcal{H}_{\infty}$ values for all $255$ possible selections of $\tilde{S}$ for $9$-bus (top-left), $14$-bus (top-right), $39$-bus (bottom-left), and $57$-bus (bottom-right) test systems. The selections leading to the instability have been excluded.}\label{fig:5}
\end{figure}

\begin{figure}[ht]
\centering

\subfloat{\includegraphics[keepaspectratio=true,scale=0.35]{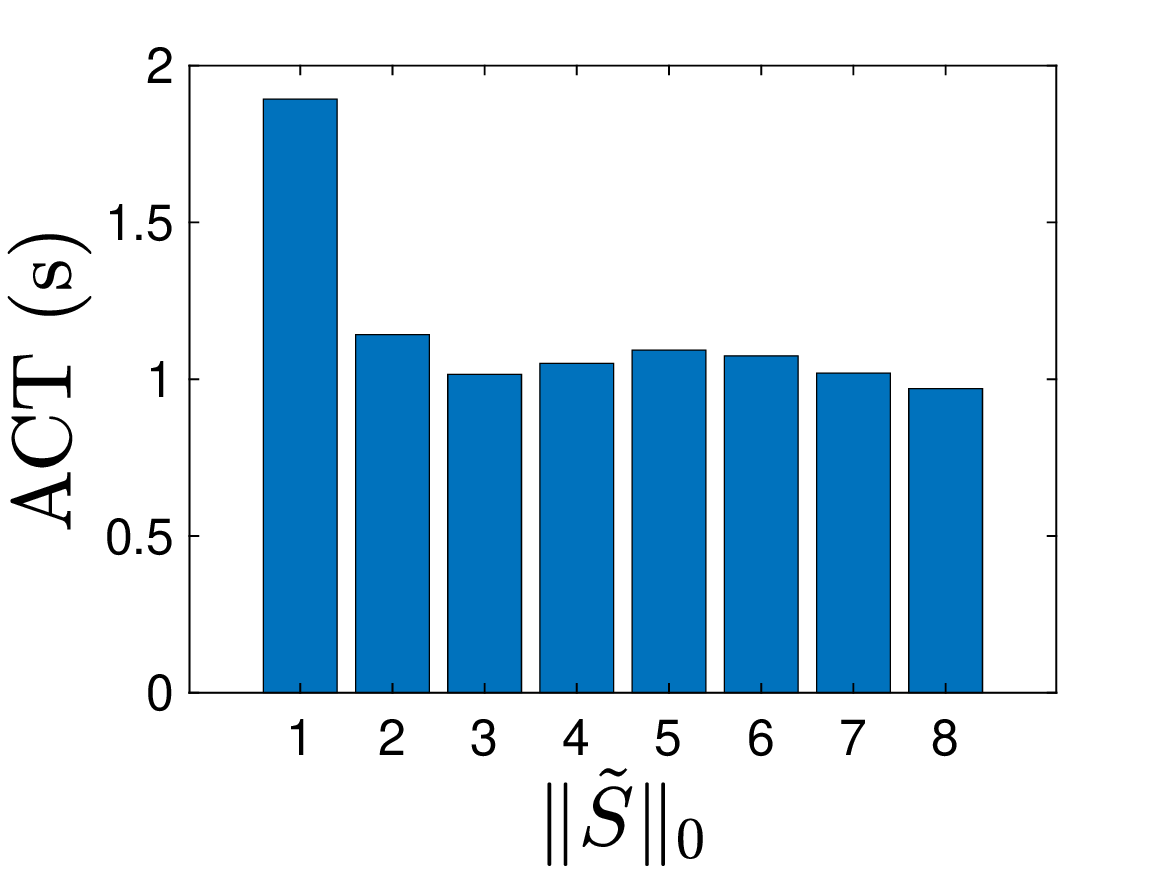}}{}{}
\subfloat{\includegraphics[keepaspectratio=true,scale=0.35]{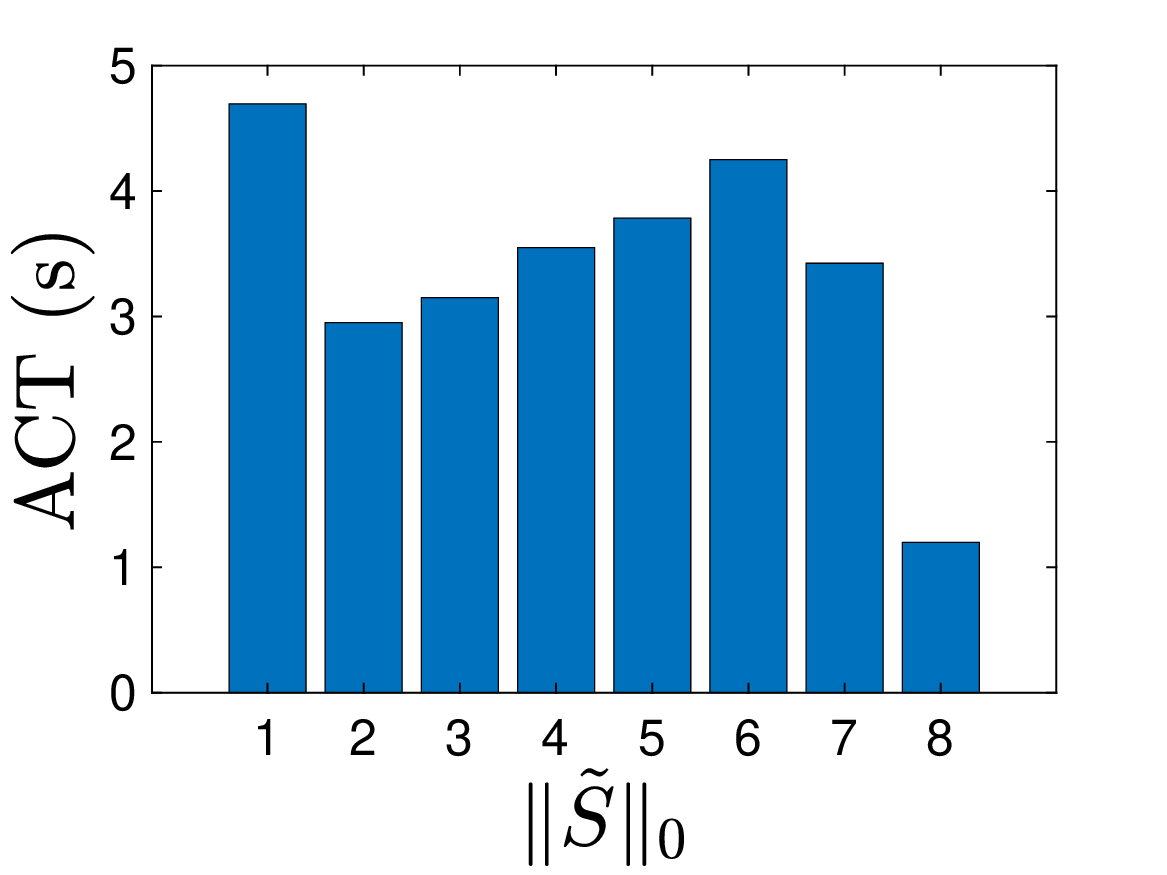}}{}{}
\\
\subfloat{\includegraphics[keepaspectratio=true,scale=0.35]{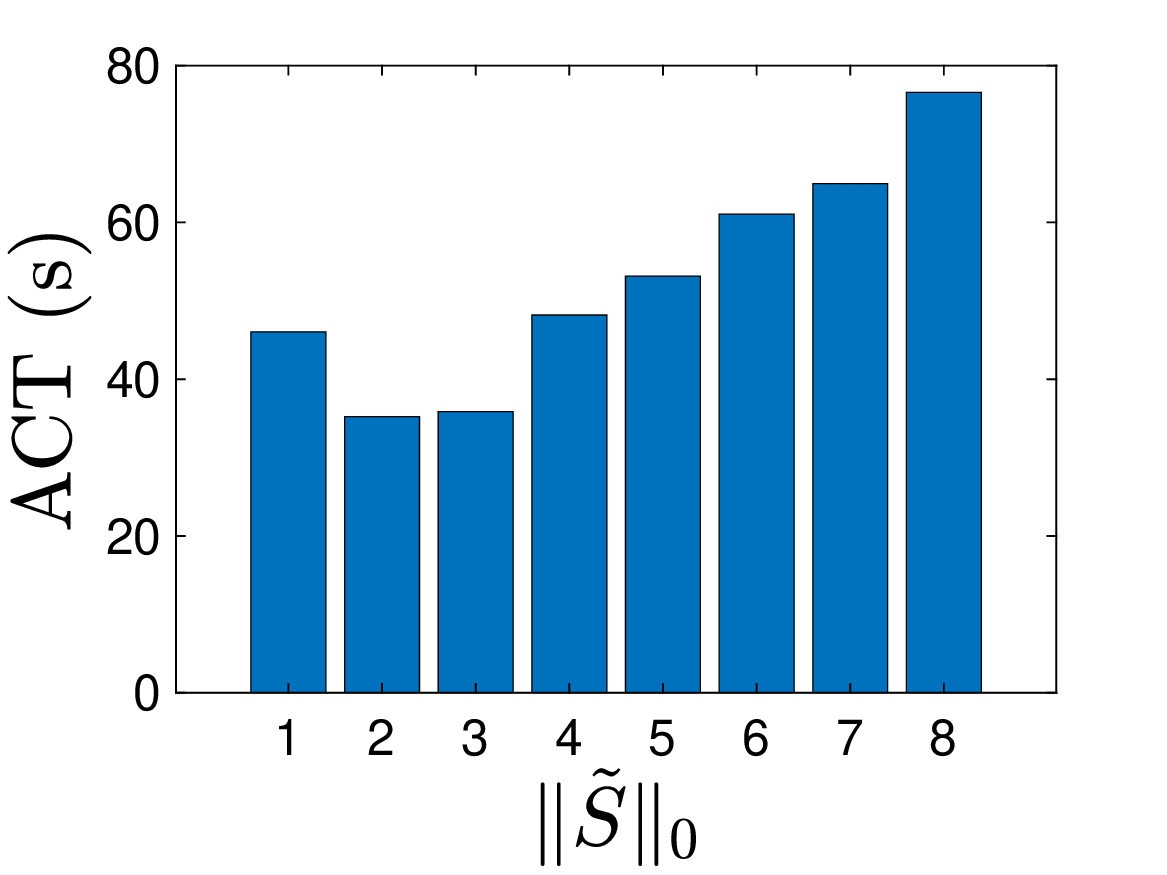}}{}{}
\subfloat{\includegraphics[keepaspectratio=true,scale=0.35]{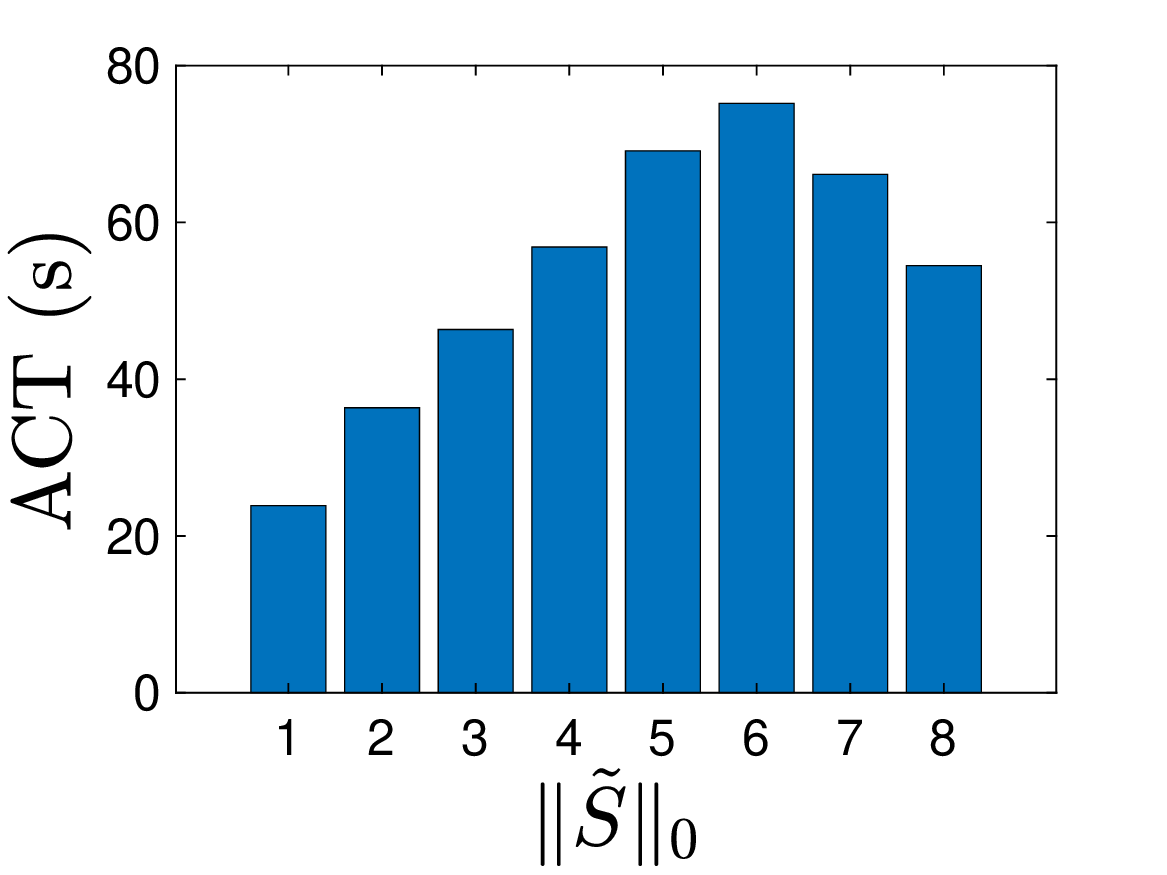}}{}{}

\caption{(Section \ref{Sec:B}) The average computational time (ACT) (s) versus $\|\tilde{S}\|_0$ for all $255$ possible selections of $\tilde{S}$ for $9$-bus (top-left), $14$-bus (top-right), $39$-bus (bottom-left), and $57$-bus (bottom-right) test systems.}\label{fig:CT}
\end{figure}

Defining the following block matrix notations for $i,j \in \{1,\dots,N\}$:
    \begin{align*}
        \hat{u} &= \begin{bmatrix}
            \hat{u}_1^T & \cdots & \hat{u}_{N}^T
        \end{bmatrix}^T,~\hat{u}_i = \begin{bmatrix} u_i & u_{i+N} \end{bmatrix}^T,~
        \hat{x} = \begin{bmatrix}
            \hat{x}_1^T & \cdots & \hat{x}_{N}^T
        \end{bmatrix}^T,\\
        \hat{x}_j &= \begin{bmatrix} x_j & x_{j+N} & x_{j+2N} & x_{j+3N} \end{bmatrix}^T,\\
        \hat{F}_{i,j} &= 
        \begin{bmatrix}
            F_{i,j} & F_{i,j+N} & F_{i,j+2N} & F_{i,j+3N}\\
            F_{i+N,j} & F_{i+N,j+N} & F_{i+N,j+2N} & F_{i+N,j+3N}
        \end{bmatrix},
    \end{align*}
we get $\hat{u} = \hat{F} \hat{x}$. Then, we can define the \textit{block-sparsity} structure of $\hat{F}$ as follows:
\begin{align*}
    \hat{S}_{i,j} &= \begin{cases} 0,
& \mathrm{if}~\hat{F}_{i,j} = 0,\\1
& \mathrm{if}~\hat{F}_{i,j} \neq 0. 
    \end{cases}
\end{align*}
Based on the definition of this block-sparsity structure, the decentralized sparsity structure $S = \mathbf{1}_{2 \times 4} \otimes I_N$ would attain the block-sparsity structure $\hat{S} = I_N$. It is noteworthy that since $\hat{S}_{i,i} = 1$ requires $\hat{F}_{i,i} \neq 0$, then a total number of $(2^{2 \times 4}-1)^N = 255^N$ sparsity structures $S$ can lead to the block-sparsity structure $\hat{S} = I_N$. In this sense, the sparsity structure $S = \mathbf{1}_{2 \times 4} \otimes I_N$ is the densest decentralized sparsity structure. Moreover, there exist a total number of $(2 \times 4)^N = 8^N$ sparsest decentralized sparsity structures (for each $i \in \{1,\dots,N\}$, $\hat{F}_{i,i}$ has only $1$ non-zero element). Although investigating the $255^N$ possibilities leading to $\hat{S} = I_N$ is computationally impossible, we may consider the following subset of those sparsity structures:
\begin{align}
    S = \tilde{S} \otimes I_N,
\end{align}
where $\tilde{S} \in \{0,1\}^{2 \times 4}\backslash \{0\}$. Note that in this specific subset, the sparsity structure of $\hat{F}$ would be $I_N \otimes \tilde{S}$. Fig. \ref{fig:5} illustrates the $\mathcal{H}_{\infty}$ values for all $255$ possible selections of $\tilde{S}$. As Fig. \ref{fig:5} shows, for the IEEE test systems, the corresponding optimal selections of $\tilde{S}$ are as follows:
\begin{align*}
    \tilde{S}^{\mathrm{opt}}_{9-\mathrm{bus}} &= \begin{bmatrix}
        0 & 1 & 1 & 1\\
        1 & 1 & 1 & 1
    \end{bmatrix},~ \tilde{S}^{\mathrm{opt}}_{14-\mathrm{bus}} = \begin{bmatrix}
        0 & 1 & 1 & 1\\
        1 & 1 & 0 & 1
    \end{bmatrix},~
\tilde{S}^{\mathrm{opt}}_{39-\mathrm{bus}} = \begin{bmatrix}
        1 & 0 & 1 & 0\\
        0 & 1 & 1 & 1
    \end{bmatrix},\\
    \tilde{S}^{\mathrm{opt}}_{57-\mathrm{bus}} &= \begin{bmatrix}
        1 & 1 & 1 & 1\\
        1 & 1 & 1 & 1
    \end{bmatrix}.
\end{align*}
with the $\mathcal{H}_{\infty}$ values of $3.2729$, $7.4251$, $9.6947$, and $23.6180$, respectively. Comparing these values with the corresponding values reflected in Tab. \ref{table:NL}, we observe that except for the $57$-bus test system, imposing the sparsity structure on $F$ can help the non-Lyapunov solver to obtain a better solution in terms of $\mathcal{H}_{\infty}$ optimality. Note that ideally the best solution should be associated with the densest sparsity structure, i.e., $\tilde{S} = \mathbf{1}_{2 \times 4}$, however, due to the sub-optimal nature of the solutions of non-Lyapunov solver in practice (highlighted by Remark \ref{R1}), it may not be the case. Nevertheless, it suggests that exploring the sparsity-constrained solutions may be beneficial to us in practice via potentially detecting better optimal solutions in terms of $\mathcal{H}_{\infty}$ optimality. Fig. \ref{fig:CT} visualizes the relationship between the sparsity level of $\tilde{S}$ and the average computational time (s). We observe that by increasing the size of the test system, computing the sparsest solutions \textcolor{black}{takes} less time compared to the densest solutions. We may interpret it as the positive effect of imposing the sparsity structure on computational time improvement for the larger test systems in which we deal with a larger search space for solutions. In other words, imposing the sparsity structure significantly reduces the search space in the case of large test systems.

\subsection{Output matrix selection} \label{Sec:C}

\subsubsection{The additionally-accessed algebraic states}
In the previous section, we chose the output matrix $C_y$ as $C_y = \begin{bmatrix} I_{n_d} & 0 \end{bmatrix}$. In other words, we assumed that we only have access to the information of dynamic states, i.e., $x_d = \begin{bmatrix} \delta^T & \omega^T & E'^T & T_M^T \end{bmatrix}^T$ which includes the generators' internal states. Here, we select a relatively general output matrix $C_y$ as $C_y = \begin{bmatrix} I_{6N} & 0 \end{bmatrix}$ by additionally taking into account the $2N$ first elements of algebraic states, i.e., $a = \begin{bmatrix} P_G^T & Q_G^T \end{bmatrix}^T$ which includes the generators' supplied power states.

Considering the following decentralized sparsity structure:
\begin{align*}
    S &= \begin{bmatrix}
       I_{N} & I_{N} & I_{N} & I_{N} & I_{N} & I_{N}\\
       I_{N} & I_{N} & I_{N} & I_{N} & I_{N} & I_{N}
    \end{bmatrix} = \mathbf{1}_{2 \times 6} \otimes I_{N},
\end{align*}
we obtain the results reflected in Tab. \ref{table:NL3}. 

\begin{table}[t]
\centering
\caption{(Section \ref{Sec:C}) The $\mathcal{H}_{\infty}$ values and computational times for NLA corresponding to SSOF $\mathcal{H}_{\infty}$ controllers with different choices of Decentralized (Dec.) $S$: $(n_u,n_y) = (2N,4N)$ and $(n_u,n_y) = (2N,6N)$, $C_y = I_{n_y,n_x}$, $D_y = 0.1I_{n_y,n_w}$, and $B_h = 0.1I_{n_x}$ for the IEEE test systems.}
\label{table:NL3}
\begin{tabular}{|c|c|c|c|}
\hline
\textrm{Sparsity Structure} & $\|T_{z\tilde{w}}(s)\|_{\mathcal{H}_{\infty}}$ & Computational Time & $(n_u,n_y)$ \\
\hline
Dec. $9$-bus & $3.2784$ & $4.11$ s & $(6,12)$ \\
\hline
Dec. $9$-bus & \cellcolor{lightgray}$3.1946$ & \cellcolor{lightgray}$2.68$ s & $(6,18)$ \\
\hline
Dec. $14$-bus & $7.4274$ & $1.78$ s & $(10,20)$ \\
\hline
Dec. $14$-bus & \cellcolor{lightgray}$7.4273$ & \cellcolor{lightgray}$1.65$ s & $(10,30)$\\
\hline
Dec. $39$-bus & $9.8467$ & $82.38$ s & $(20,40)$\\
\hline
Dec. $39$-bus & \cellcolor{lightgray}$9.4424$ & \cellcolor{lightgray}$32.17$ s & $(20,60)$\\
\hline
Dec. $57$-bus & $23.6180$ & $56.93$ s & $(14,28)$\\
\hline
Dec. $57$-bus & \cellcolor{lightgray}$23.5797$ & \cellcolor{lightgray}$21.21$ s & $(14,42)$\\
\hline
\end{tabular}
\end{table}

Tab. \ref{table:NL3} demonstrates that utilizing the information of generators' supplied power states, i.e., $2N$ extra state measurements, both $\mathcal{H}_{\infty}$ norm values and computational times can significantly be improved.

\subsubsection{The partially-accessed dynamic states}
We can consider the scenario in which, we want to optimally select and place $n_y = p$ states out of $n_d = 4N$ dynamic states in the $\mathcal{H}_{\infty}$ sense. \textcolor{black}{To this end}, there exist $\frac{n_d !}{n_y ! (n_d - n_y) !}$ possibilities in the case of $S = \mathbf{1}_{n_u \times n_y}$. Note that according to the existence of permutation matrix transformations detailed later on, we do not take into account the $n_y !$ permutations of each selection when $S = \mathbf{1}_{n_u \times n_y}$ holds. Then, denoting the positions of non-zero elements of $C_y$ as $(1,j_1)$ to $(n_y,j_{n_y})$ at which each element of $C_y$ is set to $1$, without loss of generality, we can assume that the ascending order $j_1 < \dots < j_{n_y}$ holds for the case of $S = \mathbf{1}_{n_u \times n_y}$. Assuming that $F$ and $F^{\pi}$ denote the solutions corresponding to $(j_1,\dots,j_{n_y})$ and $(j_{\pi(1)},\dots,j_{\pi(n_y)})$, respectively where $\pi: \{1,\dots,n_y\} \longrightarrow \{1,\dots,n_y\}$ represents a permutation mapping, we have 
\begin{align*}
    F^{\pi} &= F P_{\pi}^T,~P_{\pi}^T = \begin{bmatrix}
        \mathbf{e}_{\pi(1)} & \cdots & \mathbf{e}_{\pi(n_y)}
    \end{bmatrix}.
\end{align*}

\begin{figure}[ht]
    \centering

\subfloat{\includegraphics[keepaspectratio=true,scale=0.35]{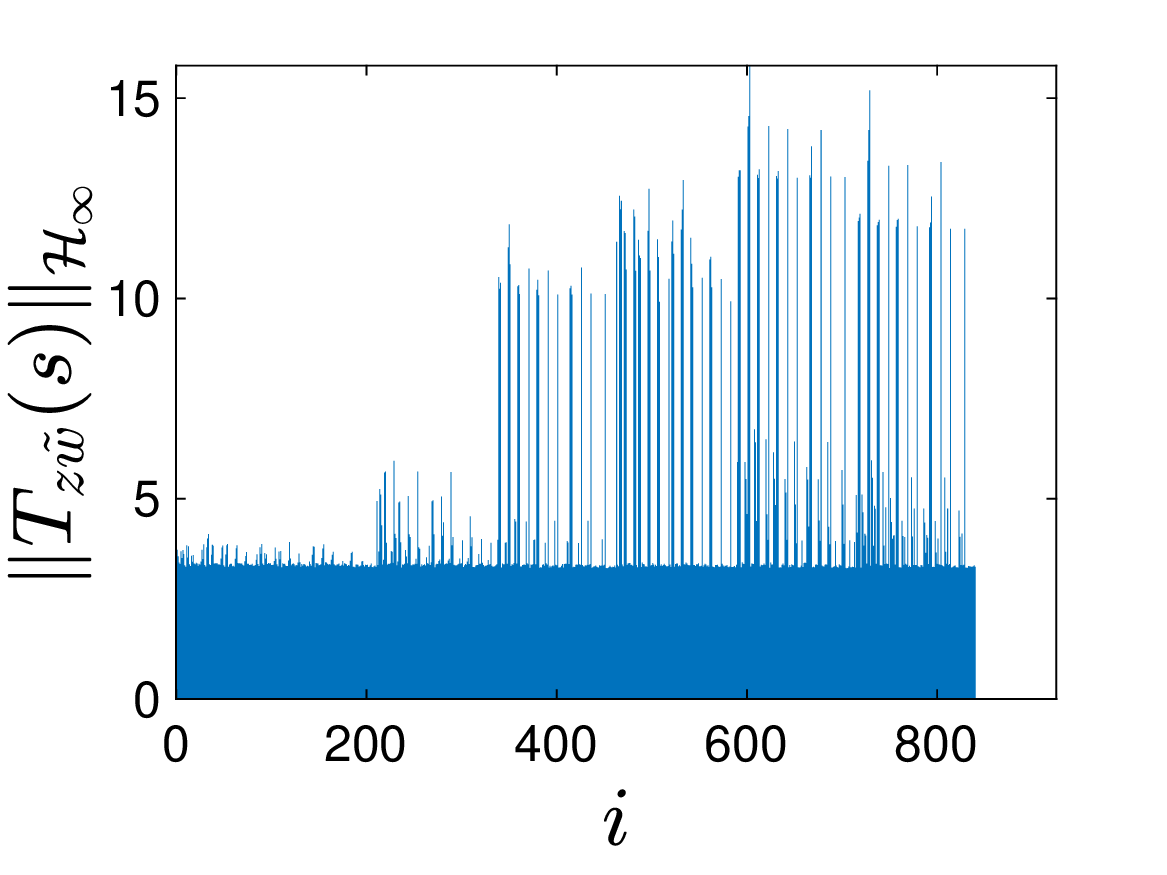}}{}{}
\subfloat{\includegraphics[keepaspectratio=true,scale=0.35]{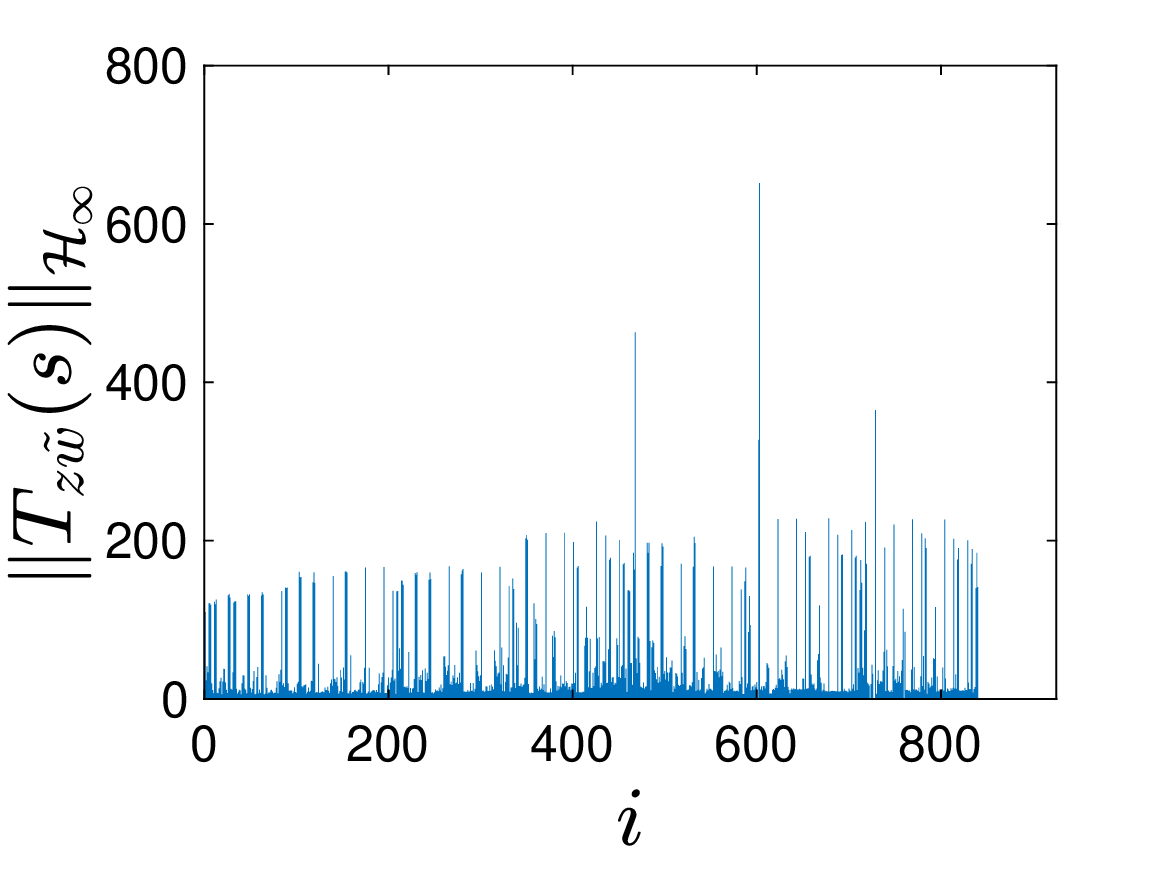}}{}{}

    \caption{(Section \ref{Sec:C}) The $\mathcal{H}_{\infty}$ values for all $\frac{n_d !}{n_y ! (n_d - n_y) !}$ possible selections of $C_y$ for $9$-bus test system: Centralized, $S = \mathbf{1}_{n_u \times n_y}$ (left), and Decentralized, $S = I_{n_u \times n_y}$ (right). The selections leading to the instability have been excluded.}
    \label{fig:my_label_CySel}
\end{figure}

However, for the case of $S = I_{n_u \times n_y}$, we need to take into account all the permutations as the decentralized sparsity structure on the controller is not preserved under the permutation matrix transformation. Then, the total number of possible selections in such a scenario would be $\frac{n_d !}{(n_d - n_y) !}$. Since investigating all $\frac{n_d !}{(n_d - n_y) !}$ possibilities is highly time-consuming, similar to the case of $S = \mathbf{1}_{n_u \times n_y}$, we only investigate the $\frac{n_d !}{n_y ! (n_d - n_y) !}$ possibilities in ascending order. Setting $n_y = p = 2N$, Fig. \ref{fig:my_label_CySel} depicts the $\mathcal{H}_{\infty}$ values for all $\frac{n_d !}{n_y ! (n_d - n_y) !}$ possible selections of $C_y$ for $9$-bus test system including both centralized $S = \mathbf{1}_{n_u \times n_y}$ and decentralized $S = I_{n_u \times n_y}$ scenarios. As Fig. \ref{fig:my_label_CySel} demonstrates, the optimal selection for the centralized scenario is associated with $i = 453$, i.e., $(j_1,\dots,j_6) = (1,6,8,9,10,12)$ and the corresponding $\mathcal{H}_{\infty}$ value is $3.2579$. Also, the optimal selection for the decentralized scenario is associated with $i = 19$, i.e., $(j_1,\dots,j_6) =(1,2,3,4,8,9)$ and the corresponding $\mathcal{H}_{\infty}$ value is $5.6785$. Fig. \ref{fig:my_label_CySel} interestingly shows that the selections after $i = 840$, i.e., the selections with $j_1 > 3$ cannot stabilize the system. In other words, if none of $(x_1,x_2,x_3)$ is selected, there is no stabilizing $F$. The physical interpretation of such an observation is that the state information of $\delta$ of at least one of the generators is required to stabilize the system.

\begin{figure}[ht]
    \centering

\subfloat{\includegraphics[keepaspectratio=true,scale=0.35]{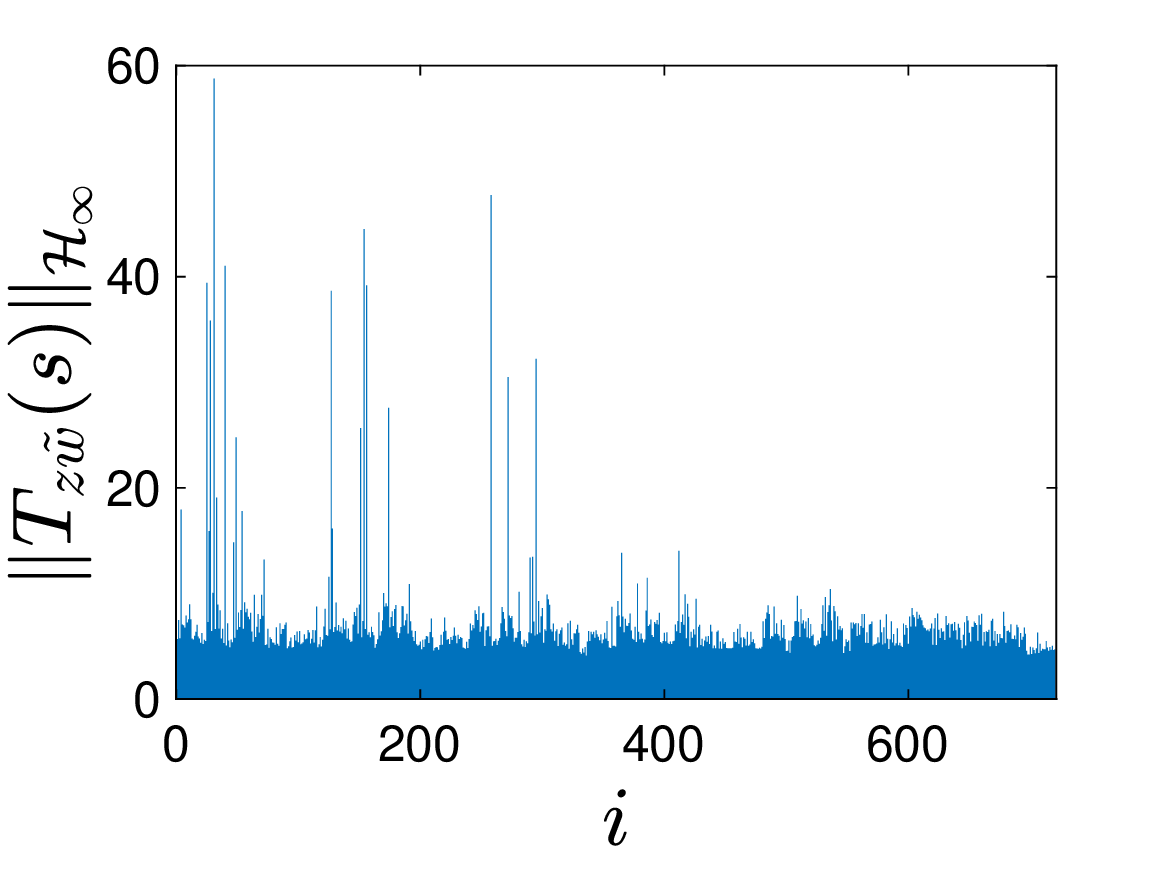}}{}{}
\subfloat{\includegraphics[keepaspectratio=true,scale=0.35]{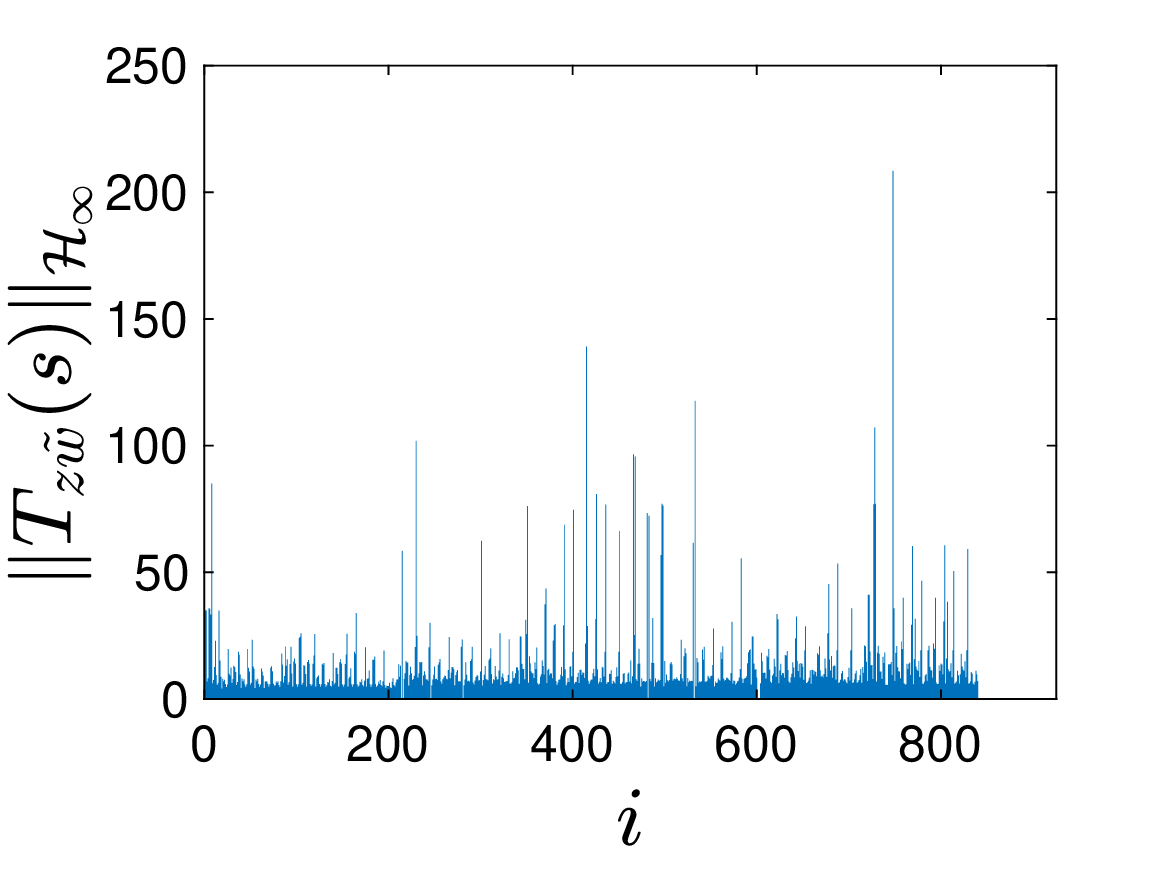}}{}{}

    \caption{(Section \ref{Sec:C}) The $\mathcal{H}_{\infty}$ values for the $n_y !$ permutations of the optimal selection of $C_y$ among the $\frac{n_d !}{n_y ! (n_d - n_y) !}$ possibilities in ascending order for $9$-bus test system, Decentralized, $S = I_{n_u \times n_y}$ (left), the $\mathcal{H}_{\infty}$ values for all $\frac{n_d !}{n_y ! (n_d - n_y) !}$ possible selections of $C_y$ subject to the permutation order associated with the optimal selection for $9$-bus test system, Decentralized, $S = I_{n_u \times n_y}$ (right (the selections leading to the instability have been excluded)).}
    \label{fig:my_label_CySelDec}
\end{figure}

In Fig. \ref{fig:my_label_CySelDec}, for $9$-bus test system and $S = I_{n_u \times n_y}$, we explore the $n_y !$ permutations of the optimal selection among the $\frac{n_d !}{n_y ! (n_d - n_y) !}$ possibilities in ascending order. Also, we explore all $\frac{n_d !}{n_y ! (n_d - n_y) !}$ possible selections of $C_y$ subject to the permutation order associated with the optimal selection. Interestingly, both explorations suggest $(j_1,\dots,j_6) =(3,8,9,4,2,1)$ with $\mathcal{H}_{\infty}$ value of $4.1145$ as the optimal placement for $C_y$. Then, we realize that exploring the $n_y !$ permutations of the optimal selection for $C_y$, i.e., $720$ possibilities associated with $6$ optimally selected states $\{1,2,3,4,8,9\}$, significantly improves the $\mathcal{H}_{\infty}$ value from $5.6785$ to $4.1145$.

\subsection{\textcolor{black}{Dense estimator design}} \label{Sec:D}

In this section, we design a dense $\mathcal{H}_{\infty}$ estimator via setting $S = \mathbf{1}_{n_x \times n_y}$. Considering the IEEE test systems, we utilize a convex SDP similar to the one proposed in \cite{nadeem2022robust} and Procedure \ref{alg:two} to obtain the Lyapunov-based and NLA dense $\mathcal{H}_{\infty}$ estimators, respectively. \textcolor{black}{In order to make a fair comparison between the Lyapunov-based approach and NLA, we slightly modify the convex SDP proposed by \cite{nadeem2022robust} provided that $\Delta f$ is $\mathcal{L}_2$-norm bounded (instead of Lipschitz boundedness assumption utilized in \cite{nadeem2022robust}). Then, we utilize a modified version of \eqref{LyaFunn} as follows:}
\begin{align*}
 \textcolor{black}{\dot{V}(e) + e^T e - \lambda \tilde{w}^T \tilde{w} < 0.}
\end{align*}

\textcolor{black}{Tab. \ref{table:LNL-Estimation} lists the $\mathcal{H}_{\infty}$ values and computational times for the Lyapunov-based approach and NLA corresponding to the dense $\mathcal{H}_{\infty}$ estimator with $S = \mathbf{1}_{n_x \times n_y}$ and $B_{\Delta f} = 0.1I_{n_x}$ for the IEEE test systems. As Tab. \ref{table:LNL-Estimation} indicates, NLA proposes better solutions in terms of $\mathcal{H}_{\infty}$ values compared to its Lyapunov counterpart. However, due to the extremely large number of variables in NLA (e.g., $12480$ for the $57$-bus test system) and the simpler (sparser) structure of the LMIs in the Lyapunov-based approach, NLA is highly time-consuming, in the case of estimator design, unlike the controller design case.}

\textcolor{black}{Tab. \ref{table:LNL-Estimationu} represents the $\mathcal{H}_{\infty}$ values and computational times for the Lyapunov-based approach and NLA corresponding to the dense $\mathcal{H}_{\infty}$ estimator with $S = \mathbf{1}_{n_x \times n_y}$, $B_{\Delta f} = 0.1I_{n_x}$, $C_y = I_{n_y,n_x}$, and $n_y = 2N$ for the IEEE test systems. Comparing the results of Tab. \ref{table:LNL-Estimationu} with the results of Tab. \ref{table:LNL-Estimation} reveals the fact that the scalability of NLA can significantly be improved for the case of having less number of output measurements (notice that the scalability superiority of the Lyapunov-based approach compared to NLA gets reversed). For instance, for the $57$-bus test system, reducing the $n_y$ from $80$ to $14$, the computational time reduces from more than $8$ hours to less than $1$ minute. Then, minimizing the number of output measurements can play a significant role in terms of scalability improvement while attaining a reasonable $\mathcal{H}_{\infty}$ performance degradation. As another observation, we realize that the $\mathcal{H}_{\infty}$ performance of the Lyapunov-based approach can significantly be improved for the case of having \textcolor{black}{a smaller} number of output measurements. For instance, except for the $14$-bus test system, the Lyapunov-based approach attains the same $\mathcal{H}_{\infty}$ performance as NLA. Although the computational time is also improved, it is not significant compared to the scalability improvement in the case of NLA.} 

\begin{table}[t]
\centering
\caption{\textcolor{black}{(Section \ref{Sec:D}) The $\mathcal{H}_{\infty}$ values and computational times for the Lyapunov-based approach and NLA corresponding to the Dense $\mathcal{H}_{\infty}$ estimator with $S = \mathbf{1}_{n_x \times n_y}$ and $B_{\Delta f} = 0.1I_{n_x}$ for the IEEE test systems.}}
\label{table:LNL-Estimation}
\begin{tabular}{|c|c|c|c|}
\hline
\textrm{Approach} & $\|T_{z\tilde{w}}(s)\|_{\mathcal{H}_{\infty}}$ & Computational Time & $(n_x,n_y)$ \\
\hline
non-Lyap $9$-bus & \cellcolor{lightgray}$1.5100$ & $4.76$ s & (36,14) \\
\hline
Lyap $9$-bus & \cellcolor{lightgray}$1.5100$ & \cellcolor{lightgray} $2.26$ s & $(36,14)$ \\
\hline
non-Lyap $14$-bus & \cellcolor{lightgray}$1.4140$ &  $44.30$ s & $(58,28)$ \\
\hline
Lyap $14$-bus & $2.6656$ & \cellcolor{lightgray} $7.58$ s & $(58,28)$ \\
\hline
non-Lyap $39$-bus & \cellcolor{lightgray}$3.1129$ & $6841.34$ s & $(138,58)$ \\
\hline
Lyap $39$-bus & $4.7038$ & \cellcolor{lightgray} $943.74$ s & $(138,58)$ \\
\hline
non-Lyap $57$-bus & \cellcolor{lightgray}$2.1155$ & $29639.22$ s & $(156,80)$ \\
\hline
Lyap $57$-bus & $3.6180$ & \cellcolor{lightgray}$4113.78$ s & $(156,80)$ \\
\hline
\end{tabular}
\end{table}

\begin{table}[t]
\centering
\caption{\textcolor{black}{(Section \ref{Sec:D}) The $\mathcal{H}_{\infty}$ values and computational times for the Lyapunov-based approach and NLA corresponding to the Dense $\mathcal{H}_{\infty}$ estimator with $S = \mathbf{1}_{n_x \times n_y}$, $B_{\Delta f} = 0.1I_{n_x}$, $C_y = I_{n_y,n_x}$, and $n_y = 2N$ for the IEEE test systems.}}
\label{table:LNL-Estimationu}
\begin{tabular}{|c|c|c|c|}
\hline
\textrm{Approach} & $\|T_{z\tilde{w}}(s)\|_{\mathcal{H}_{\infty}}$ & Computational Time & $(n_x,n_y)$ \\
\hline
non-Lyap $9$-bus & \cellcolor{lightgray}$3.0921$ & \cellcolor{lightgray}$0.85$ s & (36,6) \\
\hline
Lyap $9$-bus & \cellcolor{lightgray}$3.0921$ &  $6.06$ s & $(36,6)$ \\
\hline
non-Lyap $14$-bus & \cellcolor{lightgray}$3.2893$ & \cellcolor{lightgray}$1.95$ s & $(58,10)$ \\
\hline
Lyap $14$-bus & $4.3332$ & $7.25$ s & $(58,10)$ \\
\hline
non-Lyap $39$-bus & \cellcolor{lightgray}$2.6402$ & \cellcolor{lightgray}$45.25$ s & $(138,20)$ \\
\hline
Lyap $39$-bus & \cellcolor{lightgray}$2.6402$ & $556.77$ s & $(138,20)$ \\
\hline
non-Lyap $57$-bus & \cellcolor{lightgray}$17.5631$ & \cellcolor{lightgray}$17.28$ s & $(156,14)$ \\
\hline
Lyap $57$-bus & \cellcolor{lightgray}$17.5631$ & $1337.05$ s & $(156,14)$ \\
\hline
\end{tabular}
\end{table}

\subsection{Structured estimator design} \label{Sec:E}

In this section, we design a structured $\mathcal{H}_{\infty}$ estimator via imposing the sparsity structure $S$ on the estimator. Considering the IEEE test systems, we utilize Procedure \ref{alg:two} to obtain NLA structured $\mathcal{H}_{\infty}$ estimators.

Tab. \ref{table:NLEstimator} summarizes the $\mathcal{H}_{\infty}$ values and computational times for NLA corresponding to the dense $\mathcal{H}_{\infty}$ estimator with $S = \mathbf{1}_{n_x \times n_y}$ and the structured $\mathcal{H}_{\infty}$ estimator with $S = I_{n_x,n_y}$ under $B_{\Delta f} = 0.1I_{n_x}$ for the IEEE test systems. It shows that imposing the sparsity structure on $L$ can significantly expedite the estimator computation at the cost of $\mathcal{H}_{\infty}$ performance degradation. Such an observation is due to the fact that imposing the sparsity structure on $L$ significantly reduces the number of variables (e.g., the reduction from $12480$ to $80$ for the $57$-bus test system leading to the reduction from more than $8$ hours to $5$ minutes). Comparing the results of Tab. \ref{table:NLEstimator} with the results of Tab. \ref{table:LNL-Estimation}, suggests that for the large-scale power systems, utilizing the structured $\mathcal{H}_{\infty}$ estimator design (obtained by NLA) can significantly save computational time compared to the dense $\mathcal{H}_{\infty}$ estimator design (obtained by the Lyapunov-based approach) while degrading the $\mathcal{H}_{\infty}$ by some extent. Moreover, such performance degradation can potentially be minimized by optimally choosing the output matrix $C_y$.

\begin{table}[t]
\centering
\caption{(Section \ref{Sec:E}) The $\mathcal{H}_{\infty}$ values and computational times for NLA corresponding to structured $\mathcal{H}_{\infty}$ estimators with different choices of $S$: Dense ($S = \mathbf{1}_{n_x \times n_y}$) and Structured ($S = I_{n_x,n_y}$) under $B_{\Delta f} = 0.1I_{n_x}$ for the IEEE test systems.}
\label{table:NLEstimator}
\begin{tabular}{|c|c|c|c|}
\hline
\textrm{Sparsity Structure} & $\|T_{z\tilde{w}}(s)\|_{\mathcal{H}_{\infty}}$ & Computational Time & $(n_x,n_y)$ \\
\hline
Dense $9$-bus & \cellcolor{lightgray}$1.5100$ & $4.76$ s & $(36,14)$\\
\hline
Structured $9$-bus & $3.4120$ & \cellcolor{lightgray}$4.63$ s & $(36,14)$\\
\hline
Dense $14$-bus & \cellcolor{lightgray}$1.4140$ & $44.30$ s & $(58,28)$\\
\hline
Structured $14$-bus & $8.7075$ & \cellcolor{lightgray}$11.63$ s & $(58,28)$\\
\hline
Dense $39$-bus & \cellcolor{lightgray}$3.1129$ & $6841.34$ s & $(138,58)$\\
\hline
Structured $39$-bus & $13.3505$ & \cellcolor{lightgray}$223.11$ s & $(138,58)$\\
\hline
Dense $57$-bus & \cellcolor{lightgray}$2.1155$ & $29639.22$ s & $(156,80)$\\
\hline
Structured $57$-bus & $5.5907$ & \cellcolor{lightgray}$299.10$ s & $(156,80)$\\
\hline
\end{tabular}
\end{table}

\subsection{Estimator performance under transient conditions} \label{Sec:F}

In this section, we discuss the performance of the estimator in estimating all the states of the power system
under various transient conditions and noisy measurements. Notice that for all the case studies here, we assume that phasor measurement units (PMUs) are already placed optimally in the power system and thus, making it completely observable as given in \cite{ChakrabartiITPWRS2008,ManousakisITPWRS2013,nadeem2022dynamic}. It is noteworthy that the proposed estimator has the following key advantages as compared to the current literature on power system dynamic state estimation (DSE):
\begin{itemize}
    \item The proposed estimator in this study can simultaneously estimate both dynamic states (states of generators) and algebraic states (states of the network such as voltages and currents). In the current literature on power system DSE, they are usually estimated separately \cite{ZhangITSE2014, RouhaniITSG2018} because of the complexity of handling the complete power system NDAE models. Few studies have been carried out to estimate them simultaneously \cite{nadeem2022robust, NugrohoITCNS2022}. However, they use Lyapunov-based approaches which are much more difficult to solve for a large-scale power system model as discussed in this paper.
    \item The proposed estimator does not require any statistical properties of the disturbance/noise and can provide accurate estimation results as long as the disturbance is norm-bounded. It can also seamlessly handle the situation when the real-time control inputs are not known to the estimator and only steady-state values are given.
    \item The presented estimator also only requires a few measurements from PMUs placed optimally such that the whole system is observable as compared to the literature where it is commonly required that all the generator buses need to be equipped with PMUs \cite{ZhaoITPWRS2020}. 
\end{itemize}
For all the case studies, the estimator dynamics \eqref{eq:obsr_dynamics} are initialized from random initial conditions having $20\%$ maximum deviation from the steady-state values of the power system. With that in mind, we consider the following various scenarios:

\subsubsection{Estimation under disturbance in load demand and renewable}
\begin{figure}[ht]
\centering

\subfloat{\includegraphics[keepaspectratio=true,scale=0.7]{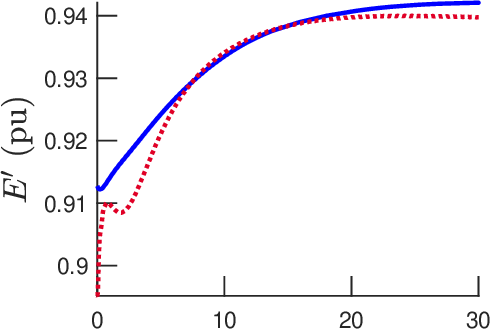}}{}{}
\subfloat{\includegraphics[keepaspectratio=true,scale=0.7]{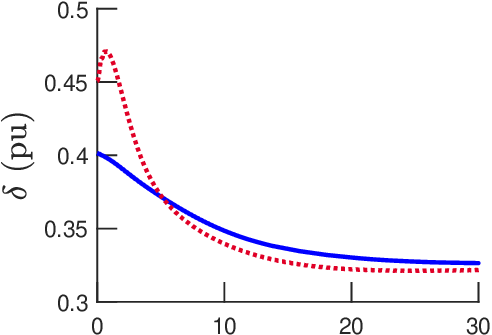}}{}{}
\\
\subfloat{\includegraphics[keepaspectratio=true,scale=0.7]{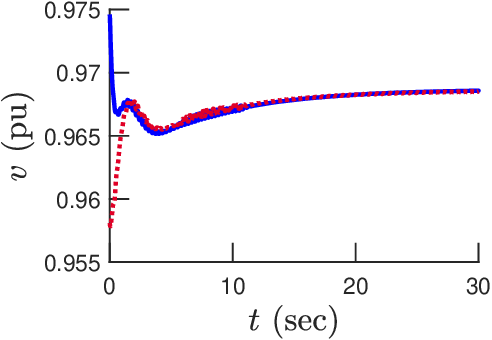}}{}{}
\subfloat{\includegraphics[keepaspectratio=true,scale=0.7]{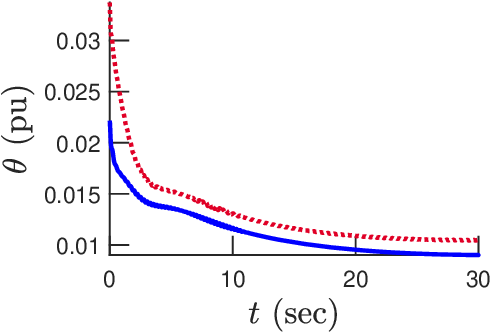}}{}{}

\caption{(Section \ref{Sec:F}) Estimation results for $57$-bus test system under load and renewable disturbance; Generator $1$ transient voltage and rotor angle and Bus $7$ voltage and angle.}\label{fig:est case 1 57}
\end{figure}

\begin{figure}[ht]
\centering

\subfloat{\includegraphics[keepaspectratio=true,scale=0.7]{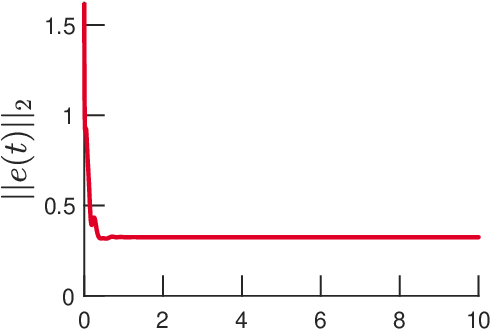}}{}{}
\subfloat{\includegraphics[keepaspectratio=true,scale=0.7]{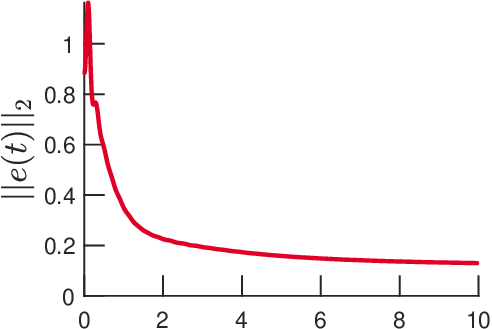}}{}{}
\\
\subfloat{\includegraphics[keepaspectratio=true,scale=0.7]{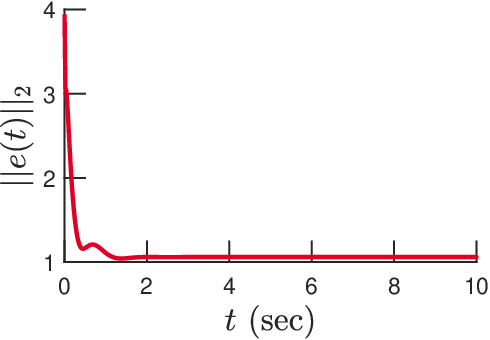}}{}{}
\subfloat{\includegraphics[keepaspectratio=true,scale=0.7]{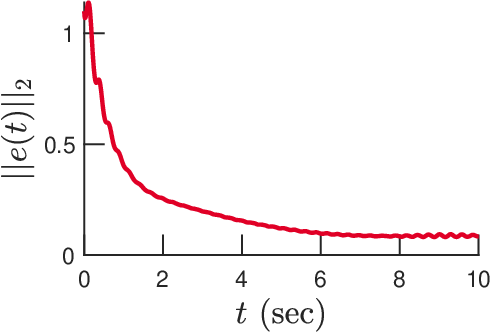}}{}{}

\caption{(Section \ref{Sec:F}) Estimation error norm for $9$-bus, $14$-bus, $39$-bus, and $57$-bus test systems under disturbance in load demand and renewable power generation.}\label{fig:est case 1}
\end{figure}

Here, we consider that the minute or hour ahead prediction (or steady-state values) of load $P_L$, $Q_L$ and renewable generation $P_R$, $Q_R$ are known (this is realistic as system operators obtain and publish these quantities regularly---see\cite{CAISO}) while the disturbance/uncertainty in them is not known to the estimator. To that end, we add disturbance in load and renewable as discussed in Section \ref{Sec:A}, and then we estimate the state of the power system using both dense and structured estimators (with only steady-state values of load and renewable known to them) as designed in the previous sections. The results are shown in Fig. \ref{fig:est case 1 57} where we see that the proposed estimator can successfully estimate all the states of the power system with appropriate accuracy. This can be verified from the error norm also given in Fig. \ref{fig:est case 1} where we observe that it asymptotically converges to zero for all test networks.

\subsubsection{Estimation under non-Gaussian measurement noise}
\begin{figure}[ht]
\centering

\subfloat{\includegraphics[keepaspectratio=true,scale=0.7]{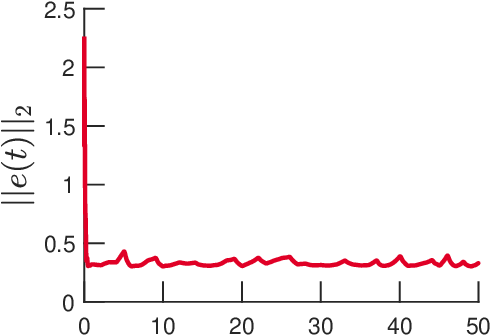}}{}{}
\subfloat{\includegraphics[keepaspectratio=true,scale=0.7]{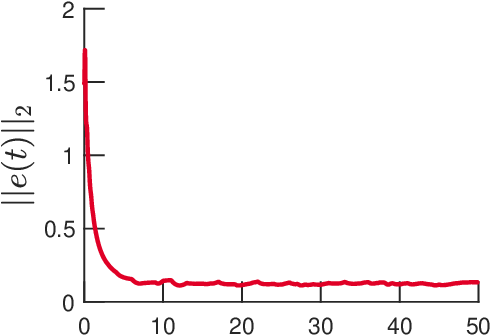}}{}{}
\\
\subfloat{\includegraphics[keepaspectratio=true,scale=0.7]{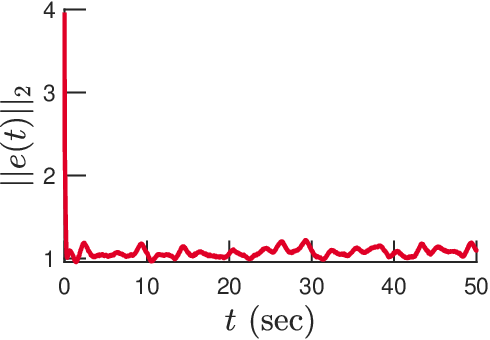}}{}{}
\subfloat{\includegraphics[keepaspectratio=true,scale=0.7]{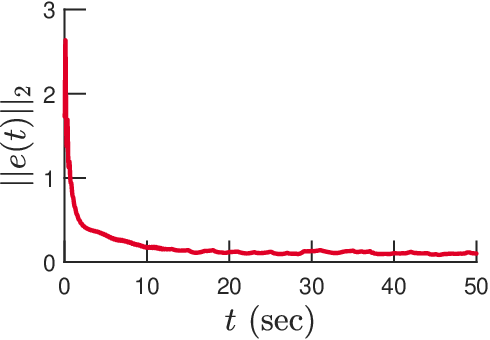}}{}{}

\caption{(Section \ref{Sec:F}) Estimation error norm for $9$-bus (top-left), $14$-bus (top-right), $39$-bus (bottom-left), and $57$-bus (bottom-right) test systems under non-Gaussian measurement noise.}\label{fig:est case 2}
\end{figure}

Here, we consider the scenario when the measurement $y$ received by the estimator contains some non-Gaussian noise. Notice that it is important to check the performance of the estimator under non-Gaussian noise because the measurement noise is usually non-Gaussian and assuming it to be Gaussian is a serious simplification as discussed in \cite{WangITPD2018}. We want to mention that the vintage Kalman-based estimators (such as EKF, UKF, etc) that are commonly used in power system DSE cannot handle non-Gaussian noise as they require some statistical properties of the noise to work. This is one of the main advantages of the proposed estimator as it can handle any type of bounded noise because of the robust $\mathcal{H}_\infty$-notion used to design it. To that end, Cauchy noise has been generated as $w_{mi} = a+b(\pi(R-0.5))$, where $a =0$, $b= 1\times 10^{-5}$ and $R$ is a random variable inside $(0,1)$ and has been added to the PMU measurements. 
The estimation results are shown in Fig. \ref{fig:est case 2} where we see that the proposed estimator can still accurately estimate all the states of the power system.
\subsubsection{Estimation under unknown control inputs}
\begin{figure}[ht]
\centering

\subfloat{\includegraphics[keepaspectratio=true,scale=0.7]{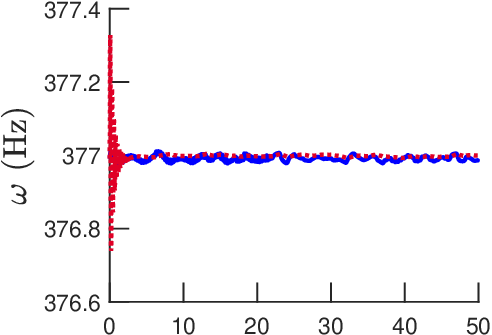}}{}{}
\subfloat{\includegraphics[keepaspectratio=true,scale=0.7]{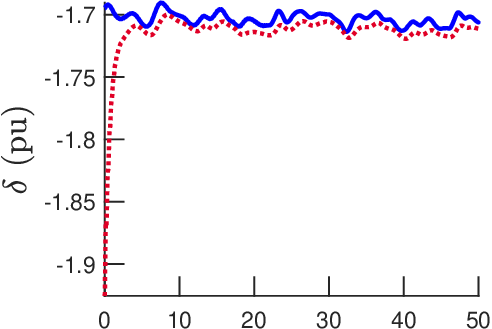}}{}{}
\\
\subfloat{\includegraphics[keepaspectratio=true,scale=0.7]{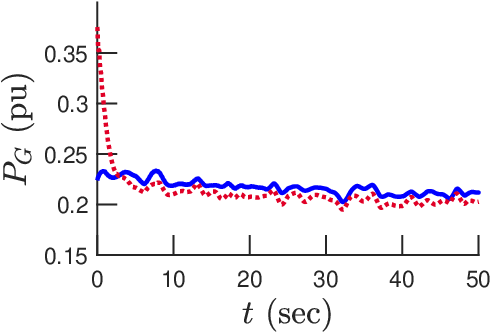}}{}{}
\subfloat{\includegraphics[keepaspectratio=true,scale=0.7]{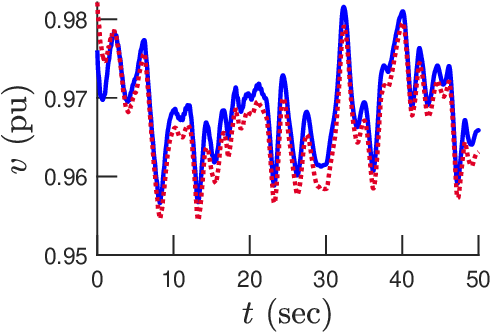}}{}{}

\caption{(Section \ref{Sec:F}) Estimation results for $57$-bus test system under unknown inputs; Generator $4$ speed, rotor angle, generated power, and terminal bus voltage.}\label{fig:est unknown inputs states}
\end{figure}

\begin{figure}[ht]
\centering
\subfloat{\includegraphics[keepaspectratio=true,scale=0.7]{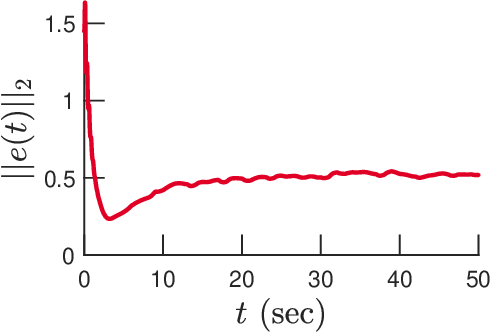}}{}{}
\subfloat{\includegraphics[keepaspectratio=true,scale=0.7]{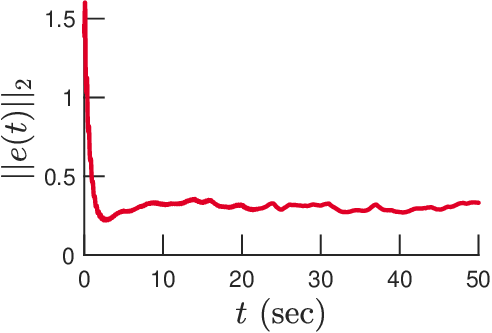}}{}{}

\caption{(Section \ref{Sec:F}) Estimation error norm for $14$-bus and $57$-bus test systems under unknown control inputs.}\label{fig:case_low_}
\end{figure}

Here, we consider the scenario when the real-time control input $u(t)$ is not available to the estimator and only steady-state values of control input $u$ are known. Notice that this is important to consider because, in the brushless excitation system of synchronous generators, it is difficult to measure field current and voltage in real-time \cite{AnagnostouITPWRS2018}. Thus, assuming that the real-time information of control inputs is known to the estimator is unrealistic. To that end only steady-state values of $u(t)$ are supplied to the estimator dynamics \eqref{eq:obsr_dynamics} and state estimation is performed. The results are shown in Fig. \ref{fig:est unknown inputs states} where we observe that the proposed estimator can still provide appropriate estimation results. This can also be verified from Fig. \ref{fig:case_low_} where we see that the trajectory of error norm between actual $x$ and estimated states $\hat{x}$ successfully converges to zero.

\section{Concluding Remarks}\label{Con}

In this paper, we \textcolor{black}{employ a celebrated non-Lyapunov approach (NLA) from the control theory literature} to design scaled-up structured $\mathcal{H}_{\infty}$ controllers and estimators. Through extensive numerical experiments on the IEEE test systems, we empirically validate that such NLA significantly improves the scalability and sensitivity simultaneously compared to its Lyapunov-based counterparts. Moreover, in the case of controller design, it can handle the cases in which we deal with partially-accessed states and noisy measurement outputs which are more realistic \textcolor{black}{compared to} the fully-accessed states and noiseless measurement states considered by the Lyapunov-based approach. Also, in the case of estimator design, the proposed estimator can deal with load demand and renewable disturbances, non-Gaussian measurement noise, and unknown control inputs. The \textcolor{black}{employed} NLA in this paper can readily be applied to various robust controller/estimator algorithms in power systems to overcome the computational scalability and computational sensitivity issues arising from the Lyapunov-based LMI formulations. \textcolor{black}{Nonetheless, NLA also has some limitations. One limitation is that NLA becomes impractical without a linear time-invariant (LTI) model. In other words, one should have an LTI model or derive an approximate LTI model before employing NLA. As another limitation, NLA cannot directly handle the continuous-time LTI models with delay while it can handle the discrete-time LTI models with delay. Furthermore, depending on the stability properties of the system, NLA can fail to stabilize the closed-loop system and may need an appropriate selection of an initialization for its built-in non-convex non-smooth optimization algorithm. Fortunately, NLA can successfully stabilize the closed-loop system in the case of power systems considered in this paper (by selecting $0$ as an initialization). Thus, in the current paper, we do not face the last expressed limitation of NLA.}

The answers to the itemized questions in Section \ref{sec:Example} (\textit{Q1}--\textit{Q5}) are as follows:
\begin{itemize}
    \item \textcolor{black}{\textit{A1}: The extensive numerical simulations (Tabs. \ref{table:LNL} and \ref{table:LNL-Estimationu}) reveal that the structured $\mathcal{H}_{\infty}$ designs can be computed via NLA in an extremely reduced computation time compared to the Lyapunov-based approach while attaining the same $\mathcal{H}_{\infty}$ performance. For instance, for the $57$-bus test system, NLA controller and estimator are $47.17$ times and $76.38$ times faster than the Lyapunov-based counterparts.}
    
    \item \textcolor{black}{\textit{A2}: For the dense controller design, as $\frac{n_u}{n_x}$ becomes larger, the scalability quality of NLA deteriorates. Similarly, as $\frac{n_y}{n_x}$ increases, the scalability quality of NLA degrades (as illustrated by Tab. \ref{table:LNL-Estimation}). As Tab. \ref{table:LNL-Estimation} shows, the superiority of NLA over the Lyapunov-based counterpart has been reversed due to the large value of $\frac{n_y}{n_x}$. In Tab. \ref{table:LNL-Estimationu}, the small value of $\frac{n_y}{n_x}$ ensures the superiority of NLA over the Lyapunov-based counterpart. An intuitive justification for such an observation could be that the smaller the number of variables, the smaller the search space for the solutions.}
    
    \item \textcolor{black}{\textit{A3}: Unlike the Lyapunov-based approach, NLA can be utilized to propose $\mathcal{H}_{\infty}$ controller designs subject to the imposed structural/sparsity constraints (e.g., distributed and decentralized) with partially-accessed states in the presence of noisy measurement outputs (Tab. \ref{table:NL} and Fig. \ref{fig:case_high_}). Tab. \ref{table:NL} reflects a fundamental trade-off between the $\mathcal{H}_{\infty}$ performance and the controller sparsity. Defining the block-sparsity structure and comparing the results demonstrated by Fig. \ref{fig:5} and Tab. \ref{table:NL}, we realize that imposing the more sparse structures may enable us to obtain the more sparse structured designs with better $\mathcal{H}_{\infty}$ performance. A possible justification behind such a counterintuitive observation could be that the smaller the number of nonzero elements (variables), the smaller the search space for the solutions. Similarly, NLA can be utilized to propose structured $\mathcal{H}_{\infty}$ estimator designs. Tab. \ref{table:NLEstimator} demonstrates that by imposing the sparsity structure to the estimator, computational time can significantly be improved at the cost of extra $\mathcal{H}_{\infty}$ performance degradation.} 
    
    \item \textit{A4}: Tab. \ref{table:NL3} depicts that by taking advantage of the information of additional algebraic states (generators' supplied power states) in addition to the dynamics states (generators' internal states), both $\mathcal{H}_{\infty}$ performance and computational time can significantly be improved. Also, Figs. \ref{fig:my_label_CySel} and \ref{fig:my_label_CySelDec} visualize the process of optimal selection/placement of $n_y = p$ states out of $n_d = 4N$ dynamic states in the $\mathcal{H}_{\infty}$ sense. As an interesting observation, we realize that the state information of $\delta$ of at least one of the generators is required to stabilize the system.
    
    \item \textcolor{black}{\textit{A5}: Figs. \ref{fig:est case 1 57}, \ref{fig:est case 1}, \ref{fig:est case 2}, \ref{fig:est unknown inputs states}, and \ref{fig:case_low_} highlight that NLA can be utilized to propose estimators capable of effectively dealing with load demand and renewable disturbances, non-Gaussian measurement noise, and unknown control inputs. The proposed estimator in this study can simultaneously estimate both dynamic states (states of generators) and algebraic states (states of the network such as voltages and currents). In the current literature on power system DSE, they are usually estimated separately because of the complexity of handling the complete power system NDAE models. The proposed estimator does not require any statistical properties of the disturbance/noise and can provide accurate estimation results as long as the disturbance is norm-bounded. The presented estimator also only requires a few measurements from PMUs placed optimally such that the whole system is observable as compared to the literature where it is commonly required that all the generator buses need to be equipped with PMUs.}
\end{itemize}

\bibliographystyle{elsarticle-num} 
\bibliography{cas-refs}

\appendix

\section{The Lyapunov-based approach to design dense controllers} \label{ApendxA}

The dense static state feedback (DSSF) $\mathcal{H}_{\infty}$ controller proposed by \cite{nadeem2023robust} is obtained as follows:
\begin{align} \label{LyCo}
& \boxed{F = H(X E^T + E^{\perp} W)^{-1}},
\end{align}
for which the following convex SDP is solved accordingly:
\begin{align*}
\underset{\lambda,H,X,W}{\min}&~\lambda\\
\mathrm{s.t.}&~X \succ 0, \lambda > 0,\begin{bmatrix} \mathbf{He}(A(XE^T+E^{\perp}W) + BH) & \ast & \ast \\ \hat{B}_w^T & -\lambda I & \ast \\ C(X E^T + E^{\perp} W)+DH & \hat{D}_w & -I \end{bmatrix} \prec 0.
\end{align*}

\section{The Lyapunov-based approach to design dense estimators} \label{ApendxB}

The dense $\mathcal{H}_{\infty}$ estimator proposed by \cite{nadeem2022robust} is obtained as follows:
\begin{align} \label{LyEs}
& \boxed{L = (N(X E + E^{T \perp} Y)^{-1})^T},
\end{align}
for which the following convex SDP is solved accordingly:
\begin{align*}
\underset{\epsilon,\lambda,N,X,Y}{\min}&~\lambda\\
\mathrm{s.t.}&~X \succ 0, \epsilon > 0, \lambda > 0,\begin{bmatrix} \Omega & \ast & \ast \\ X E + E^{T \perp} Y & -\epsilon I & \ast \\ B_w^T(X E + E^{T \perp} Y) & 0 & -\lambda I \end{bmatrix} \prec 0.
\end{align*}
where $\Omega = \mathbf{He}(A^T(XE+E^{T \perp}Y) + C_y^T N)+ \epsilon \alpha^2 I$.

\end{document}

%% file: preamble.tex
\usepackage{color,amsmath}
\usepackage{graphicx}
\usepackage[table]{xcolor}

\usepackage[linesnumbered,boxed,commentsnumbered,ruled,vlined,longend]{algorithm2e}
\SetAlgorithmName{Procedure}

\usepackage{amsmath}    %For theorems
\usepackage{bm}
\usepackage{epstopdf}
\usepackage{comment}
\usepackage{amssymb}
\usepackage{url}
\usepackage{multirow}
\usepackage{hhline}
\usepackage{booktabs}
\usepackage{comment}

\usepackage{tikz}
\usetikzlibrary{shapes.geometric, arrows}

\tikzstyle{startstop} = [rectangle, rounded corners, 
minimum width=3cm, 
minimum height=1cm,
text centered, 
draw=black, 
fill=red!30]

\tikzstyle{io} = [trapezium, 
trapezium stretches=true, % A later addition
trapezium left angle=70, 
trapezium right angle=110, 
minimum width=3cm, 
minimum height=1cm, text centered, 
draw=black, fill=blue!30]

\tikzstyle{process} = [rectangle, 
minimum width=3cm, 
minimum height=1cm, 
text centered, 
text width=3cm, 
draw=black, 
fill=orange!30]

\tikzstyle{arrow} = [thick,->,>=stealth]

\makeatother
\DeclareMathAlphabet\mathbfcal{OMS}{cmsy}{b}{n}

\newtheorem{mypbm}{Problem}

\newtheorem{myrem}{Remark}



\usepackage{stackengine}

\newcommand{\cs}{{\cal s}}

\allowdisplaybreaks[4]
\usepackage[colorlinks = true,
linkcolor = blue,
urlcolor  = blue,
citecolor = blue,
anchorcolor = blue]{hyperref}

\newcommand{\mr}[1]{\mathrm{#1}}
\usepackage[framemethod=TikZ]{mdframed}
\mdfdefinestyle{MyFrame}{%
	linecolor=black,
	outerlinewidth=1.25pt,
	roundcorner=1.25pt,
	innerrightmargin=5pt,
	innerleftmargin=5pt,}

\usepackage[noabbrev]{cleveref}

\usepackage{mathtools}

\DeclarePairedDelimiter\abs{\lvert}{\rvert}%
\DeclarePairedDelimiter\norm{\lVert}{\rVert}%

\makeatletter
\let\oldabs\abs
\def\abs{\@ifstar{\oldabs}{\oldabs*}}
\let\oldnorm\norm
\def\norm{\@ifstar{\oldnorm}{\oldnorm*}}
\makeatother

\usepackage[english]{babel}
\usepackage[utf8]{inputenc}
\usepackage[super]{nth}

\usepackage{float}
\usepackage[caption = false]{subfig}

\usepackage{array}
\usepackage{threeparttable}

\usepackage[english]{babel}
\usepackage[utf8]{inputenc}
\usepackage[super]{nth}

\usepackage{pifont}% http://ctan.org/pkg/pifont